# Room-temperature sub-100 nm Néel-type skyrmions in non-stoichiometric van der Waals ferromagnet $Fe_{3-x}GaTe_2$ with ultrafast laser writability

**Authors:** Zefang Li[1#], Huai Zhang[2#], Guanqi Li[3#], Jiangteng Guo[1], Qingping Wang[4], Ying Deng[1], Yue Hu[1], Xuange Hu[1], Can Liu[1], Minghui Qin[2], Xi Shen[5], Richeng Yu[5], Xingsen Gao[2], Zhimin Liao[6], Junming Liu[2,7], Zhipeng Hou[2*], Yimei Zhu[8*] & Xuewen Fu[1,9*]

**Affiliations:**

[1] Ultrafast Electron Microscopy Laboratory, The MOE Key Laboratory of Weak-Light Nonlinear Photonics, School of Physics, Nankai University, Tianjin 300071, China

[2] Guangdong Provincial Key Laboratory of Optical Information Materials and Technology, Institute for Advanced Materials, South China Academy of Advanced Optoelectronics, South China Normal University, Guangzhou, 510006, China

[3] School of Integrated Circuits, Guangdong University of Technology, Guangzhou 510006, China

[4] School of Physics and Electronic and Electrical Engineering, Aba Teachers University Wenchuan

[5] Beijing National Laboratory for Condensed Matter Physics, Institute of Physics, Chinese Academy of Sciences, Beijing, 100190 China

[6] State Key Laboratory for Mesoscopic Physics and Frontiers Science Center for Nano-optoelectronics, School of Physics, Peking University, Beijing 100871, China

[7] Laboratory of Solid State Microstructures and Innovation Center of Advanced Microstructures, Nanjing University, Nanjing 211102, China

[8] Condensed Matter Physics and Materials Science Department, Brookhaven National Laboratory, Upton, New York 11973, United States

[9] School of Materials Science and Engineering, Smart Sensing Interdisciplinary Science Center, Nankai University, Tianjin 300350, China

*Corresponding authors:

houzp@m.scnu.edu.cn (Z.H.); zhu@bnl.gov (Y.Z.); xwfu@nankai.edu.cn (X.F.).

#These authors contribute equally to this work.

# ABSTRACT

Realizing room-temperature magnetic skyrmions in two-dimensional van der Waals ferromagnets offers unparalleled prospects for future spintronic applications. However, due to the intrinsic spin fluctuations that suppress atomic long-range magnetic order and the inherent inversion crystal symmetry that excludes the presence of the Dzyaloshinskii-Moriya interaction, achieving room-temperature skyrmions in 2D magnets remains a formidable challenge. In this study, we target room-temperature 2D magnet $Fe_3GaTe_2$ and unveil that the introduction of iron-deficient into this compound enables spatial inversion symmetry breaking, thus inducing a significant Dzyaloshinskii-Moriya interaction that brings about room-temperature Néel-type skyrmions with unprecedentedly small size. To further enhance the practical applications of this finding, we employ a homemade in-situ optical Lorentz transmission electron microscopy to demonstrate ultrafast writing of skyrmions in $Fe_{3-x}GaTe_2$ using a single femtosecond laser pulse. Our results manifest the $Fe_{3-x}GaTe_2$ as a promising building block for realizing skyrmion-based magneto-optical functionalities.

## INTRODUCTION

Magnetic skyrmions, which are topological swirling spin configurations stabilized by Dzyaloshinskii–Moriya interaction (DMI)[1,2], have garnered significant interest over the past decade because of their nanometric scale and exotic magnetoelectronic properties[3], such as the topological Hall effect (THE)[4], skyrmion Hall effect[5], and ultra-low current density for motion[6,7], making them ideal information carriers for future high-density and fast-speed data storage[8], quantum and neuromorphic computation[9,10]. Since the existence of skyrmion crystal was first verified in helimagnet MnSi[11], various skyrmion-host three-dimensional (3D) bulk materials have been discovered with abundant magnetic and electronic features[12,13]. Compared to the 3D bulk ferromagnets, two-dimensional (2D) van der Waals (vdW) ferromagnets have inherent superiorities for practical applications in spintronic devices due to their unique atomic layered structure[14,15], such as long-range magnetic order down to atomic thickness[16,17], wide flexibility for stacking artificial heterostructures[18], high sensitivity to external field perturbations[19], and possible compatibility with modern integrated circuit process[20]. Therefore, exploring magnetic skyrmions with small size, especially at room temperature (RT), in 2D vdW ferromagnetic materials with facile tunability have become a focal point of magnetic and topological order of matters, as well as spintronic applications[21].

To realize RT magnetic skyrmions within the isolated 2D vdW ferromagnets, DMI that induced by inversion crystal symmetry breaking and RT ferromagnetism are crucial prerequisites[22]. Nevertheless, for most of the intrinsic 2D vdW ferromagnets discovered hitherto, such as $Fe_3GeTe_2$[23], $Fe_5GeTe_2$[24], $CrGeTe_3$[25] and $CrI_3$[26] etc., on one hand, their natural centrosymmetric crystal structures exclude the DMI[27]; on the other hand, due to the strong Mermin–Wagner fluctuations that suppress the intrinsic magnetic order in the 2D limit[28], their Curie temperatures ($T_c$) are typically below RT. Therefore, on the way pursing RT magnetic skyrmions in 2D vdW ferromagnets, how to simultaneously improve $T_c$ and introduce DMI is an internationally recognized difficulty. Although several methods, such as elemental doping and substitution etc.[29,30], have shown the possibility to increase $T_c$ while breaking the centrosymmetric structure for creating DMI in intrinsic 2D vdW ferromagnets[31,32], the elemental controllability and raising the

$T_c$ above RT are still challenging[33,34]. One better choice is to directly introduce DMI into intrinsic 2D vdW ferromagnets with $T_c$ above RT. The iron-based ternary telluride $Fe_3GaTe_2$[35] and chromium-based binary telluride $CrTe_2$[36] are the only two 2D vdW intrinsic magnets that exhibit RT ferromagnetism discovered hitherto, but merely the $Fe_3GaTe_2$ exhibits both above $T_c$ and large RT perpendicular magnetic anisotropy (PMA) up to the order of magnitudes of $10^5$ J/m³ [35]. Recently, a $Fe_3GaTe_2$-based magnetic tunnel junction (MTJ) achieved a large tunnel magnetoresistance (TMR) of 85% at RT[37], highlighting the great potential of $Fe_3GaTe_2$ for developing spintronic devices. However, the intrinsic inversion symmetric crystal structure of $Fe_3GaTe_2$ possess an inherent obstacle. Hitherto, the universal strategy that enables directly breaking the inversion symmetry and achieving substantial DMI for stabilizing RT skyrmions in the $Fe_3GaTe_2$ have not been realized.

In this study, we discover that the iron deficiency in $Fe_3GaTe_2$ can lead to a pronounced displacement of the Fe atoms within the crystal structure. Based on systematic structural analysis and first-principles calculations, we find that this atomic displacement causes a transformation from the original centrosymmetric crystal structure to a non-centrosymmetric structure, resulting in a significant DMI. Combined Lorentz transmission electron microscopy (LTEM), magnetic force microscopy (MFM) and magneto-transport measurements demonstrate that the non-stoichiometric $Fe_{3-x}GaTe_2$ could accommodate Néel-type skyrmions together with a prominent topological Hall effect over a broad temperature range of 330 K to 100 K. Moreover, the size of the skyrmions decreases as the sample thickness becomes thinner, and field-free sub-100 nm skyrmions can be obtained at RT when the thickness falls below a threshold ranging from 40 to 60 nm, which are the smallest skyrmions achieved hitherto in 2D vdW magnets. More intriguingly, with the use of a homemade in-situ optical LTEM, we realize an ultrafast writing of RT skyrmions in the non-stoichiometric $Fe_{3-x}GaTe_2$ thin flakes by a single femtosecond (fs) laser pulse, which offers a possible avenue for the realization of ultrafast and energy-efficient skyrmion-based logic and memory devices.

**RESULTS AND DISCUSSIONS**

In order to control the Fe content, we systematically grew a series of $Fe_{3-x}GaTe_2$

single crystals by varying the Fe content in the raw material composition, utilizing a Te-flux method (see Methods section and Supplementary Note 1). To determine the chemical composition of the as-grown crystals, energy dispersive X-ray spectroscopy (EDX) analyses were conducted on the surfaces of $Fe_{3-x}GaTe_2$ nanoflakes (Fig. 1a) that were exfoliated and placed onto the $Si_3N_4$ membrane (see Methods). The ratio of raw materials and the corresponding final crystal composition are listed in Table S1, Supplementary Fig. S1 and Fig. 1b. We found that the Fe deficiencies always exist in these crystals, while the minimum and maximum Fe contents correspond to $Fe_{2.84\pm0.05}GaTe_2$ and $Fe_{2.96\pm0.02}GaTe_2$, respectively. This result implies the feasibility of inducing Fe deficiency in the samples. To highlight the existence of Fe vacancies, the subsequent studies were focused on the minimum Fe content sample $Fe_{2.84\pm0.05}GaTe_2$. The Raman measurements revealed a clear red-shift of the $A_1$ peak in the non-stoichiometric $Fe_{2.84\pm0.05}GaTe_2$ to 126 cm$^{-1}$, in contrast to 130 cm$^{-1}$ observed for the stoichiometric $Fe_3GaTe_2$ reported previously (Fig. 1c)[35], which can be attributed to the localized effect of Fe-vacancy defects[38]. We conducted the magnetic characterization of the $Fe_{2.84\pm0.05}GaTe_2$ crystal with the external magnetic field ($B$) applied along the out-of-plane direction. Fig. 1d displays the temperature-dependent magnetization ($M$-$T$) curve measured under a small magnetic field of 30 mT using a field-cooled protocol. By calculating the first derivative (d$M$/d$T$) of the $M$-$T$ curve, we determined the Curie temperature ($T_c$) of the $Fe_{2.84\pm0.05}GaTe_2$ crystal to be ~ 350 K, which is slightly lower than that of the stoichiometric $Fe_3GaTe_2$. Additionally, an obvious magnetization kink was observed in the $M$-$T$ curve at around 290 K, as indicated by the dashed box in Fig. 1d. This particular kink is typically regarded as a signature of rotation of the magnetic easy axis, and is often observed in skyrmion-hosting magnetic systems[23,39]. Moreover, the field-dependent magnetization curves for the $Fe_{2.84\pm0.05}GaTe_2$ sample with minimum Fe content reveal an out-of-plane easy magnetization direction at room-temperature (Fig. S2). These curves exhibit magnetic anisotropy almost identical to that of $Fe_{2.96\pm0.02}GaTe_2$ sample with high Fe content.

The crystal structure of the stoichiometric $Fe_3GaTe_2$ was conformed to have the centrosymmetric space group of $P6_3/mmc$, and can be visualized as a series of Te-$Fe_3$Ga-Te monolayers stacked along the $c$-axis (Supplementary Fig. S3). Within each monolayer, the central $Fe_{ii}$-Ga slice flanked by two adjacent $Fe_i$ slices is sandwiched between two outer Te slices, indicating a protected $c \rightarrow -c$ mirror symmetry. Regarding the non-stoichiometric $Fe_{2.84\pm0.05}GaTe_2$ in this work, we analyzed the crystal structure on a microscopic scale (high-resolution scanning transmission electron microscopy, HR-STEM) and a macroscopic scale (single-crystal X-ray diffraction, XRD). Fig. 1e presents a typical high-angle annular dark field (HAADF) image of the $Fe_{2.84\pm0.05}GaTe_2$ nanoflake along the [0001] zone axis. The image demonstrates that the $Fe_i$ atom columns (dark brown circles) are hexagonally surrounded by six Te-$Fe_{ii}$-Te-Ga atomic columns (marked with light brown and pale green balls), matching well with the standard $Fe_3GaTe_2$ crystal structure. Moreover, its lattice parameters ($a = b = 4.08$ Å) are slightly enlarged compared to those of the stoichiometric $Fe_3GaTe_2$ ($a = b = 3.99$ Å)[35]. These observations suggest that the Fe-deficiency does not cause noticeable lattice distortion in the $ab$ plane. However, the HAADF image along the [11$\bar{2}$0] zone axis (Fig. 1f) reveals that $Fe_{ii}$ atoms deviate clearly from the center position of the Te slices along the $c$-axis, which is also supported by the annular bright-field (ABF-STEM) image in Fig. S4. By referencing the center of the two Te atoms in a magnified ABF-STEM image (Fig. S5), an averaged $Fe_{ii}$ deviation is calculated as $-0.16 \pm 0.06$ Å over an area of 2 × 17 unit cells (see Supplementary Note 2). Notably, this deviation can be also identified in the bottom layer of the $Fe_{2.84\pm0.05}GaTe_2$ unit cell, and both the offset and deviation direction are the same as those observed in the top layer. To gain additional insights into the structural features of the $Fe_{2.84\pm0.05}GaTe_2$, we performed single-crystal XRD on a single crystal with dimensions of approximately 3 × 3 mm². A total of 534 symmetry-independent reflections, corresponding to Miller indices of $-5 \leq h \leq 5$, $-5 \leq k \leq 5$, $-20 \leq l \leq 21$, were collected for the structural determination and refinement. It is well known that for the centrosymmetric space group $P6_3/mmc$ of stoichiometric $Fe_3GaTe_2$, the reflection patterns, such as ($hh\bar{2h}l$) and ($000l$), are permitted only for even values of $l$ ($l = 2n$, where $n$ are integers), whereas they are forbidden for the odd values ($l = 2n + 1$) [33,40]. In the case of $Fe_{2.84\pm0.05}GaTe_2$, however, a series of weak ($11\bar{2}l$) and ($000l$) reflection patterns, such as ($11\bar{2}3$) and ($11\bar{2}5$) (see Fig. S6), were detected for $l = 2n + 1$, suggesting a substantial deviation from the original crystal structure of $Fe_3GaTe_2$. By fitting the reflections while allowing for the relaxation of $z$-positions and occupation ratio of Fe atoms, we obtained a symmetry-lowering structural model with the non-centrosymmetric space group $P3m1$ (Supplementary Note 2 and Fig. S7). The refined crystal structure revealed that the $Fe_{ii}$ atoms are deficient and located at the Wyckoff site *1c* and *1b* at $(x, y, z) = (2/3, 1/3, 0.7482)$ and $(1/3, 2/3, 0.2486)$ with a slight deviation of $Fe_{ii}$ atoms from the Ga-Ga plane, as shown in Fig. 1g. Based on the experimentally established crystal structure (Fig. S8a), we simulated the HR-STEM images along the [0001] and [11$\bar{2}$0] zone axis (Fig. S8b and S8c), which agree well with our experimental observations[41]. Moreover, we also simulated the selected area electron diffraction (SAED) along the [10$\bar{1}$0] and [11$\bar{2}$0] zone axes. The presence of odd $l$ values in the (000$l$) diffractions, evident in both simulated and experimental SAED results (Fig. 1h and Fig. S9), further confirms that the crystal structure of the $Fe_{2.84\pm0.05}GaTe_2$ belong to the non-centrosymmetric $P3m1$ space group, rather than the centrosymmetric $P6_3/mmc$ structure of $Fe_3GaTe_2$. These simulations agree well with our experimental observations, providing solid evidential support for our structural model's reliability.

In comparison to the stoichiometric $Fe_3GaTe_2$ with a centrosymmetric structure, the presence of Fe deficiency in $Fe_{2.84\pm0.05}GaTe_2$ should exert a pivotal influence on the $Fe_{ii}$ deviation for the asymmetric structure. Our refined single-crystal XRD indicates that Fe deficiency is predominantly concentrated at the $Fe_{ii}$ positions with an occupancy ratio of 0.8467. Additionally, the upper-layer $Fe_{i-a}$ sites have an occupancy ratio of 0.9688, while the under-layer $Fe_{i-b}$ sites are nearly fully occupied (Fig. 1g). As observed in the line profile of $Fe_{i-a}$ and $Fe_{i-b}$ atoms in the ABF-STEM image (Fig. S5c), it is apparent that the image intensity of $Fe_{i-a}$ above $Fe_{ii}$ is weaker than that of $Fe_{i-b}$ below $Fe_{ii}$. Since the ABF imaging intensity is generally proportional to the number of projected atoms[42], the contrast difference between $Fe_{i-a}$ and $Fe_{i-b}$ indicates asymmetric site occupations, suggesting a small quantity of Fe vacancies in the $Fe_{i-a}$ site, which is consistent with the results of single-crystal XRD. To assess the influence of $Fe_{i-a}$ and $Fe_{ii}$ vacancies on $Fe_{ii}$ deviation, we further conducted first-principles calculations involving structure relaxation under three scenarios: no vacancy, $Fe_{i-a}$ vacancy, and $Fe_{ii}$ vacancy (Supplementary Note 3). The electron density of $Fe_{3-x}Ga$ atoms is depicted in Fig. S10 to facilitate a comparison of the alterations in $Fe_{ii}$ chemical bonding: (a) The perfect Fe$_3$GaTe$_2$, with no Fe vacancy, showcases a hexagonally bonded Fe$_{ii}$-Ga plane. In this arrangement, the centrally positioned Fe$_{i-a}$ and Fe$_{i-b}$ dimers do not form direct bonds with Fe$_{ii}$ and Ga atoms. Thus, the overall chemical bonding is mirror-symmetric along the Fe$_{ii}$-Ga plane with no Fe$_{ii}$ deviation. (b) The presence of Fe$_{ii}$ vacancy induces deformation of the Ga electron density within the *ab* plane. Nevertheless, no bonding is established between the Fe$_{ii}$-Ga plane and the Fe$_i$ atoms, which remain a mirror-symmetric electron density with no Fe$_{ii}$ deviation. (c) In case of Fe$_{i-a}$ vacancy, there is additional electron-density overlapping between the lower Fe$_{i-b}$ atom and its three nearest Fe$_{ii}$ atoms, while no overlapping between Fe$_{i-b}$ and its three nearest Ga atoms. As a result, chemical bonding between the Fe$_{ii}$ and Fe$_{i-b}$ atoms induces a substantial Fe$_{ii}$ deviation, with a calculated $\delta c$ (Fe$_{ii}$ − Ga) of about −0.0554 Å, which compares favorably to the XRD result of −0.0871 Å. Furthermore, the calculated formation energy value for Fe$_i$ vacancy (2.96 eV/Fe) is higher than Fe$_{ii}$ vacancy (2.86 eV/Fe), indicating that the formation of Fe$_{ii}$ vacancies is more favorable. The above Fe vacancy model and calculated Fe$_{ii}$ deviation align well with the analysis from single-crystal XRD and ABF-STEM image. Therefore, we conclude that the asymmetric vacancy of Fe$_{i-a}$ induces a displacement of Fe$_{ii}$ atoms towards the –*c* direction, which results in the symmetry breaking of the Fe$_{2.84\pm0.05}$GaTe$_2$ crystal structure.

Due to the asymmetric crystal structure, a pronounced DMI is naturally expected in the non-stoichiometric Fe$_{2.84\pm0.05}$GaTe$_2$. In quantitative terms, the DMI vector can be expressed as

$$\mathbf{D} = D \cdot \left( \hat{\mathbf{u}}_{ij} \times \hat{\mathbf{z}} \right), \qquad (1)$$

where $D$ is the DMI constant, $\hat{\mathbf{u}}_{ij}$ represents the unit vector from Fe$_i$ atom to Fe$_{ii}$ atom, and $\hat{\mathbf{z}}$ represents the unit vector from magnetic Fe$_{ii}$ atom to heavy Te atom. For Fe$_{2.84\pm0.05}$GaTe$_2$, since its structural symmetry is broken by the Fe$_{ii}$ atom deviation, the DMI is proposed to originate from the interactions between the neighboring Fe$_i$-Fe$_{ii}$ pair and the adjacent Te atoms. As schematically shown in Fig. 1i, each Fe$_{2.84\pm0.05}$GaTe$_2$ unit cell has two distinct DMI sources: (a) the interaction between the Fe$_i$-Fe$_{ii}$ atom pair and the upper Te atom (corresponding to $\mathbf{D}_1$ vector, represented by the red arrow) and (b) the interaction between the Fe$_i$-Fe$_{ii}$ pair and the lower Te atom (corresponding to $\mathbf{D}_2$ vector, represented by the blue arrow). Equation (1) indicates that the two DMI vectors

$\mathbf{D}_1$ and $\mathbf{D}_2$ are perpendicular to the Fe$_i$-Fe$_{ii}$-Te triangle of atoms. This is consistent with the Fert-Levy DMI observed at heavy metal/ferromagnet interfaces, which can lead to the formation of Néel-type skyrmions[2]. Moreover, we found that the directions of $\mathbf{D}_1$ and $\mathbf{D}_2$ are opposite due to the opposite directions of $\hat{\mathbf{z}}$ for $\mathbf{D}_1$ and $\mathbf{D}_2$. For the centrosymmetric Fe$_3$GaTe$_2$, because the Fe$_{ii}$ atom is located at the center of two adjacent Te atoms, the upper $\mathbf{D}_1$ and lower $\mathbf{D}_2$ vectors would always cancel out with each other, resulting in an effective net DMI vector ($\mathbf{D}_{eff}$) of zero. In the case of Fe$_{2.84\pm0.05}$GaTe$_2$, however, the deviation of Fe$_{ii}$ atoms shows −0.0871 Å atom displacement towards −$c$ directions, which breaks the inversion symmetry, and thus makes the nonequal $\mathbf{D}_1$ and $\mathbf{D}_2$ yield a nonzero $\mathbf{D}_{eff}$ within each monolayer. As for the total DMI ($D_{total}$) of the unit cell, since the $\mathbf{D}_{eff}$ vectors in top and bottom layers have the same direction, the magnitude of $D_{total}$ is the sum of the two vectors. Moreover, Fig. S11b illustrates the effective net DMI vectors $D_{eff}$ viewed from [0001] zone axis, which are perpendicular to the Fe$_i$-Fe$_{ii}$-Te atom cross sections and exhibit threefold rotational symmetry within the $ab$ plane. Based on the model depicted in the aforementioned illustration, we quantitatively investigated the relationship between the Fe$_{ii}$ deviation value $\delta c$(Fe$_{ii}$ − Ga) and the DMI constant $D$ (see Supplementary Note 4 and Fig. S12 for details). For a centrosymmetric structure ($\delta c = 0$), the absence of Fe$_{ii}$ deviation yields $D = 0$ mJ/m$^2$. Conversely, a non-centrosymmetric structure ($\delta c = -0.0871$ Å), determined by single-crystal XRD, corresponds to $D = 0.91$ mJ/m$^2$. This significant DMI constant meets the crucial requirement for the formation of skyrmions[43-46], suggesting that the non-stoichiometric Fe$_{2.84\pm0.05}$GaTe$_2$ has the potential to exhibit topological magnetism.

For a magnetic system hosting topological spin configurations, the total Hall resistivity ($\rho_{xy}$) typically comprises three components[4,47]:

$$\rho_{xy}(H) = \rho_{xy}^N + \rho_{xy}^A + \rho_{xy}^T = R_0 H + S_A \rho_{xx}^2 M(H) + \rho_{xy}^T, \qquad (2)$$

where $\rho_{xy}^N = R_0 H$ is the normal Hall resistivity, $R_0$ is the normal Hall coefficient; $\rho_{xy}^A = S_A \rho_{xx}^2 M(H)$ is the anomalous Hall resistivity, $S_A$ is the scaling coefficient, $\rho_{xx}$ is the longitudinal resistivity, $M(H)$ is the magnetic field-dependent magnetization; and $\rho_{xy}^T$ is the topological Hall resistivity. Of these components, $\rho_{xy}^T$, which is driven by the local spin chirality, is widely regarded as a key transport signature of topological spin configurations. To investigate the potential existence of $\rho_{xy}^T$ in Fe$_{2.84\pm0.05}$GaTe$_2$, we fabricated a Hall device with a sample thickness of 250 nm (Fig. 2a) and conducted measurements of $\rho_{xy}$ with the external magnetic field applied along the normal direction of the device over the temperature range of 350 - 10 K, as shown in Fig. 2b. The results indicate that $\rho_{xy}$ varies in a nonlinear manner with sweeping the magnetic field, suggesting the presence of a ferromagnetic-order induced anomalous Hall effect. We subsequently fitted $\rho_{xy}$ using the $M(H)$ and $\rho_{xx}(H)$ curves in the high-field region based on Equation (2). Fig. 2c displays both the fitting and experimental $\rho_{xy}(H)$ curves at 300 K. It is clear that there is a discrepancy at the low-field region, indicating the presence of a pronounced THE component in $\rho_{xy}$. The extracted magnetic field-dependent $\rho_{xy}^T$ curves over the temperature range of 350 - 10 K are summarized in Fig. 2d. It is important to note that the maximum value of $\rho_{xy}^T$ (~ 1.13 µΩ cm) at 300 K is one order of magnitude higher than that of the material systems hosting RT skyrmions, such as Co-doped 2D magnets (Fe$_{0.5}$Co$_{0.5}$)$_5$GeTe$_2$ ( ~ 0.8 µΩ cm )[31], Kagome ferromagnets Fe$_3$Sn$_2$ (~ -0.4 µΩ cm)[47], and Ir/Fe/Co/Pt multilayers (~ 0.03 µΩ cm)[48], and is comparable to that of noncoplanar ferromagnet Cr$_5$Te$_6$ (~ 1.6 µΩ cm) at 90 K[49]. Furthermore, we investigated thickness- and magnetic field-dependent Hall resistivity $\rho_{xy}$ and topological Hall resistivity $\rho_{xy}^T$ under room temperature (Fig. S13). As the sample thickness increases, the topological Hall signals gradually strengthen and shift towards higher magnetic field. More importantly, the THE signals persist over a broad temperature range and various thickness, suggesting the existence of topological spin configurations in Fe$_{2.84\pm0.05}$GaTe$_2$.

To directly visualize the possible topological spin textures associated with the topological Hall effect, we conducted cryo-LTEM experiments on an exfoliated [0001]-oriented Fe$_{2.84\pm0.05}$GaTe$_2$ nanoflake with a thickness of approximately 100 nm, as schematically illustrated in Fig. 3a. For the cryo-LTEM measurements, we heated the sample above $T_c$ (380 K) and then cooled it to the desired temperature with a liquid nitrogen TEM holder under a 30 mT out-of-plane magnetic field. Following this field-cooling procedure, no magnetic contrast in the Lorentz phase images was observed when the electron beam was injected along the normal direction of the nanoflake at 300 K (Fig. 3b and Supplementary Fig. S14). However, if we titled the nanoflake an angle

($\theta$) of ±20° away from the horizontal plane, bubble-like domains with half-dark/bright contrast were detected. It is well known that the magnetic contrast discernible in a Lorentz phase image acquired via Fresnel imaging mode is predominantly caused by the deflection of the electron beam due to the in-plane magnetic field as it passes through a magnetic sample. Since in the Bloch-type domain walls the magnetization, or spin, direction is out-of-plane, the Lorentz deflection is thus in-plane and perpendicular to the walls, yielding clear domain walls contrast. In contrast, in the Néel-type domain walls, the magnetization direction in the walls is in-plane and the Lorentz deflection is along the wall length, resulting in no contrast. However, once the sample is tilted, this is no longer the case as an out-of-plane magnetization component is generated[50-52]. The angle-dependent contrast modulation seen in our experiments unambiguously suggests that the bubble-like domains we imaged are Néel-type. We further analyzed their in-plane spin structures using the transport-of-intensity equation (TIE)[53], as displayed in the right panel of Fig. 3b. We found that the in-plane magnetic induction was composed of a pair of conjoined clockwise and counterclockwise spin swirls, which agrees well with the calculated magnetic induction map for the Néel-type skyrmions[31,52]. On the basis of the deduced double in-plane swirls and Néel-type spin arrangements, we conclude that the observed bubble-like domains are indeed Néel-type skyrmions.

Due to the high $T_c$ of $Fe_{2.84±0.05}GaTe_2$, its magnetic phase is expected to remain stable well above room temperature. The upper panel of Fig. 3c presents the Lorentz phase images taken at a zero magnetic field over the temperature range of 300 to 340 K after the field-cooling operation (see Methods). The results indicate that the skyrmion phase remain stable up to 330 K, which represents a record-high value compared with that of the skyrmion-hosting magnetic vdW materials reported to date. However, when the temperature exceeds 330 K, the magnetic skyrmions became elongated and gradually transformed into stripe domains due to the presence of thermal fluctuations, which are proposed to be large enough to overcome the topologically protected energy barrier between skyrmions and stripe domains above 330 K. Furthermore, the magnetic field-dependent domain evolution process after the field-cooling operation was studied within a temperature range of 340 - 100 K (lower panel of Fig. 3c and Supplementary

Fig. S15). To illustrate the correlation between the magnetic states and both the magnetic field and temperature, we construct a magnetic phase diagram (Fig. 3d), in which the red region denotes that the sample hosts high-density skyrmions at the corresponding *T-B* plane, while the blue region indicates the absence of skyrmions. As displayed in Fig. 3d, high-density, field-free skyrmions can be stabilized in a wide temperature range of 100 - 330 K, demonstrating that $Fe_{2.84\pm0.05}GaTe_2$ is a promising material platform for use in spintronic devices.

We further use magnetic force microscopy (MFM), which is sensitive to the out-of-plane magnetic field in the sample, to study the effect of nanoflake thickness on the stabilization of the skyrmion phase at room temperature after field-cooling manipulation. All the nanoflakes used for the MFM measurements were freshly exfoliated from the same batch to ensure consistency (Supplementary Fig. S16). Additionally, we used a low-moment MFM tip (<10 mT) to minimize the influence of the tip's magnetic field on the domain structures during scanning. Fig. 4a illustrates a series of MFM images that record the magnetic field-dependent domain evolution processes at six typical thicknesses. Notably, the vertically arranged MFM images represent the same sample area under different out-of-plane fields, while the horizontal direction represents the variation in thickness for different samples. For the 250 nm thick sample, the skyrmions are arranged densely into honeycomb lattices with a large skyrmion size ($d_{sk}$) of approximately 180 nm at zero magnetic field. As the external magnetic field increases, there is little change in $d_{sk}$, accompanied, however, by a decrease in density with the sudden annihilation of skyrmions at 210 mT. When the sample reaches saturation magnetization, the MFM images with uniform contrast represent the fully ferromagnetic (FM) state. Varying the sample thickness could significantly affect the skyrmion size due to the change in the strength of dipole-dipole interaction[33,54]. Upon decreasing the sample thickness, the skyrmion size at a zero magnetic field reduces correspondingly, reaching a minimum skyrmion size of 87 nm when the sample thickness is decreased below 100 nm (Fig. 4b). This value is much smaller than that of the skyrmions in other skyrmion-hosting vdW magnets reported to date, such as $Cr_{1+x}Te_2$ (~ 400 nm)[34], $Fe_3GeTe_2$ (~ 250 nm)[55], $Fe_5GeTe_2$ (~ 200 nm)[24], and $(Fe_{0.5}Co_{0.5})_5GeTe_2$ (~ 150 nm)[31], as displayed in Fig. 4c (see Supplementary Fig.

S17 for detailed size distribution of field-free skyrmions in $Fe_{2.84\pm0.05}GaTe_2$ at varied temperatures, and Fig. S18 for the corresponding Micromagnetic simulations). Fig. 4d provides a comprehensive overview of the RT skyrmion phase diagram, depicting the relationship between thickness, magnetic field, and the occurrence of skyrmion states. The field-free skyrmions are stable in a broad thickness range and the highest density appears to be between 46 - 60 nm. As the thickness of $Fe_{2.84\pm0.05}GaTe_2$ nanoflake decreases to 25 nm, the stripe domains remain while the field-cooling process no longer generates skyrmions. Compared with the previously reported skyrmion-hosting 2D material $(Fe_{0.5}Co_{0.5})_5GeTe_2$[31], $Fe_{2.84\pm0.05}GaTe_2$ exhibits smaller magnetic parameters such as DMI constant $D$, saturation magnetization $Ms$, threshold of sample thickness $t$ and etc. (see Fig. S19 and Supplementary Note 5 for the determination of magnetic parameters), which contribute to the reduction in skyrmion size (see Table S3, Fig. S20 and Supplementary Note 6 for the corresponding micromagnetic simulations). Moreover, our micromagnetic simulations demonstrate the influence of in-plane magnetic field $B$ on skyrmion shape, revealing a progressive transformation from a circular to an elliptical configuration (see Fig. S21 and Supplementary Note 7).

Achieving ultrafast and energy-efficient writing of skyrmions is a critical requirement in the pursuit of their practical applications in high-speed and low-power spintronic devices. Conventional field-cooling operations are unsuitable as they involve heating the entire sample above the $T_c$ and slow cooling to the desired temperature[55-57]. In contrast, fs laser pulses can demagnetize a localized region on the micrometer scale within fs timescale followed by a picosecond (ps) thermal quenching[58-64]. This unique approach enables the creation of metastable magnetic states at extremely short timescales, offering a promising pathway for the ultrafast writing of skyrmions. Several studies have reported successful writing of skyrmions using short laser pulses in various materials, including magnetic multilayer films[60-62], FeGe[63], and $Co_9Zn_9Mn_2$[64]. To further investigate the possibility of ultrafast writing of RT skyrmions in the 2D van der Waals $Fe_{2.84\pm0.05}GaTe_2$ nanoflakes, we performed Lorentz phase microscopy measurements under fs laser excitation using our homemade in-situ optical LTEM setup that enables single-shot fs laser pulse excitation on the sample (520 nm wavelength, 50 μm focal spot size, and 300 fs pulse duration), as illustrated in Fig. 5a. We recorded the fs laser writing processes by taking snapshots of Lorentz phase images before and after the laser pulse excitation (fluence of ~ 11 mJ/cm$^2$), while the electron beam was blanked during laser pulse excitation to avoid unexpected sample damage. Upon laser pulse excitation of the initial domains, ultrafast heating with electron temperature ($T_{elec}$) above $T_c$ would temporarily melt the long-range magnetic order (Fig. 5b) on the fs time scale[59,60,63]. Subsequently, the paramagnetic state is rapidly quenched on the ps timescale due to the electron-phonon coupling, where the phonon temperature ($T_{phon}$) follows the variation of $T_{elec}$ (see Supplementary Note 8 and Fig. S22 for detailed calculations of temperature evolutions based on a two-temperature model). Such quenching process leads to a rapid decrease of the temperature to below $T_c$ and initiates the remagnetization process of the spin system. Since only skyrmion is a stable solution for the Fert-Levy DMI observed in Fe$_{2.84\pm0.05}$GaTe$_2$, it is expected that skyrmion states could be created during the thermal relaxation in the quenching process.

To demonstrate the feasibility of using a fs laser pulse to write RT skyrmions in the 2D van der Waals Fe$_{2.84\pm0.05}$GaTe$_2$ material, we conducted a step-by-step single-shot fs laser pulse excitation measurement on different initial field-polarized magnetic states, including stripe and single domains. These initial field-polarized domain structures were achieved by varying the out-of-plane magnetic field from a saturation state (±2 T), as indicated by the normalized Hall hysteresis curve in Fig. 5c. Moreover, the corresponding Lorentz phase images captured the magnetic states before and after the laser pulse excitation at six representative magnetic fields (see Supplementary Figs. S23 and S24 for more details). These images provide compelling evidence that the presence of an external assisting magnetic field is vital for the successful fs laser writing of skyrmions in the material[60-63]. At $B = 0$ mT, only stripe domains were observed after the laser excitation (fluence of ~ 11 mJ/cm$^2$). As $B$ is swept towards positive saturation, the width of the stripe domains gradually decreases. After the fs laser excitation, the magnetic states accessible by the laser pulse initially exhibit a mixed state of stripe domains and skyrmions at 11 mT and then transition to fully-formed skyrmions with the highest density at 46 mT. As $B$ is increased to 80 mT, the initial stripe domains start to fade out, and the laser-accessible skyrmions completely vanish, transforming into a single-domain state. Compared to the single domain state stabilized by a pure magnetic field ($B$ = 104 mT), the assisting magnetic field decreases significantly, suggesting that for the laser-writing operation the laser functions as an efficient field that can effectively shift the magnetic states, possessing a higher energy state in the magnetization process to a lower energy state. When the magnetic field is increased to 104 mT, the initial stripe domains transform into a single domain state, and the laser pulse with the used fluence can no longer induce a magnetic phase transition. However, as the magnetic field is decreased from the positive saturation field, the single domain state persists at a much lower magnetic field of 46 mT due to the hysteresis effect. Interestingly, we find that there exists a magnetic field window (69 mT to 46 mT) where laser-accessible skyrmions can be created from the initial single domain state, even though it is in a higher energy state than that of the skyrmions in the magnetization process. As the magnetic field continues to decrease, the behavior under laser excitation in the low-field region (35 mT to 0 mT) is similar to that at the beginning of the positive field sweep (0 mT to 35 mT). These results indicate that the laser-writing operation is independent of the initial domain structures but rather dependent on the strength of the assisting magnetic field. Furthermore, conventional zero-field cooling can only result in the formation of interconnected, relatively long stripe domains, but does not spontaneously lead to the creation of skyrmions[57]. To demonstrate the differences with in-site fs laser quenching approach, we conducted fluence-dependent laser pulse excitation without magnetic field (see Fig. S25 and Supplementary Note 9). After a single laser pulse with fluence of 1.3 mJ/cm$^2$, stripe domains show slightly domain wall movement. Upon increasing to 9.4 mJ/cm$^2$, a hybrid state with both stripes and skyrmions are formed, while at 11 mJ/cm$^2$ only stripe domains observed. This indicates that a higher fluence of fs laser can completely demagnetize the sample during the laser-writing process, regardless of the initial magnetic state.

To further understand the underlying magnetization dynamic process of the ultrafast laser writing of skyrmions in Fe$_{2.84\pm0.05}$GaTe$_2$ under external magnetic field assistance, we performed finite-element micromagnetic simulations on the subsequent magnetic structure evolution of an initial stripe domain after a fs laser pulse excitation at a certain external magnetic field based on the Landau-Lifshitz-Gilbert equation with Langevin dynamics[65-67]. We considered the following scenarios for the simulations: (i) the fs laser pulse interacts with the magnetic structures through photothermal effect; and (ii) the fs pulse heats the sample above the $T_c$ and melts the electronic spin structures, but without changing the atomic lattice. Specifically, the laser quenching-induced magnetization dynamics was achieved by relaxing the magnetic system from the laser-induced paramagnetic state under an out-of-plane magnetic field of 46 mT (see Supplementary Note 5 and Fig. S19 for more details about the simulations). As shown in Fig. 5d (see details in Supplementary Movie 1), following the excitation by the femtosecond (fs) laser pulse, the initial melted spin state (snapshot at $t_0$) rapidly evolves into numerous nanoscale spin clusters. These clusters contain topological defects, including skyrmionic and anti-skyrmionic nucleation centers (snapshot at $t_1$). This transformation occurs due to the ultrafast cooling, achieved at a quenching rate of up to $10^{12}$ K/s[58,68]. Because only skyrmion is a stable solution for the Fert-Levy DMI observed in $Fe_{2.84\pm0.05}GaTe_2$, in the further cooling process the anti-skyrmionic nucleation centers merge and annihilate with the nearby skyrmionic nucleation centers and appear less frequently until completely disappear, namely, the topological fluctuations[60,69], resulting in formation of a pure skyrmionic state (snapshot at $t_2$). The skyrmion nucleation process is complete at $t_2$, but the skyrmion size has not reached the energy equilibrium state. In order to minimize the magnetostatic energy, the domain walls move and gradually evolve into a uniform skyrmion lattice (snapshot at $t_3$). The micromagnetic simulations reproduce well the magnetic phase transitions observed in our experiments, confirming the reliability of our established relationship between the laser-writing process, magnetic field strength, and magnetic domain evolution. Such evidenced controllable ultrafast laser writing of skyrmions in 2D van der Waals magnetic materials provides opportunities for both fundamental researches and device applications towards magneto-optical control of spin topologies.

We report the discovery of a field-free sub-100 nm Néel-type skyrmion state in non-stoichiometric $Fe_{2.84\pm0.05}GaTe_2$ over a broad temperature range from 330 K to 100 K. Using HR-STEM and single-crystal XRD, we determine that the deviation of $Fe_{ii}$ atoms from the center position of the Te slices due to the asymmetric $Fe_{i-a}$ vacancies induces a transformation of the crystal structure from centrosymmetric to non-centrosymmetric, enabling the formation of skyrmions through the in-plane isotropic

DMI. LTEM (along with MFM) shows the size of the skyrmions decreases with sample thickness, and a field-free sub-100 nm skyrmion state was achieved at RT within a specific sample thickness range of 40 nm to 60 nm. Furthermore, we demonstrate that a single fs-laser pulse can rapidly generate field-free sub-100 nm skyrmions from both stripe domains and single domains. Our study demonstrates not only the non-stoichiometric $Fe_{2.84\pm0.05}GaTe_2$ to be a promising material platform for exploring magnetic skyrmions, but also the fs-laser can be a powerful tool to manipulate and control topological chiral spin textures to realize skyrmion-based high-speed logic and memory applications.

## METHODS

**Single crystal growth and structure characterization.**

Single crystals of $Fe_{3-x}GaTe_2$ were grown by self-flux method. The mixtures of Fe (99.99%), Ga (99.99%), and Te (99.99%) elements were mounted in an alumina crucible and sealed inside a quartz tube under high vacuum (~$10^{-4}$ Pa). The mixtures were firstly heated at 1150°C over a period of 24 h, then followed by slow-cooling down to 850°C for 3 weeks. Finally, excessive molten flux was centrifuged in order to separate the single crystals. Its chemical composition was determined by energy dispersive x-ray spectroscopy mapping (EDX, Bruker Nano GmbH Berlin). A comprehensive overview of the raw material composition and the final crystal composition are listed in **Supplementary Note 1** and **Table S1**. The single-crystal X-ray diffraction was carried out with four-circle diffractometer (Bruker D8 venture). And the refined crystal structure was solved and refined by using the Bruker SHELXTL Software Package. The HR-STEM images were acquired by high-resolution transmission electron microscope (HRTEM, JEOL ARM200F). Magnetometry measurements were carried out with the Quantum Design PPMS.

**LTEM and MFM measurements.**

The fresh $Fe_{2.84\pm0.05}GaTe_2$ nanoflakes utilized for Hall devices, LTEM and MFM experiments were prepared through an all-dry mechanical-transfer method within an argon-filled glovebox. In an argon-protected environment, the nanoflakes were first produced on PDMS stamp by micromechanical cleavage, and then transferred onto $Si_3N_4$ membrane or $SiO_2$/Si substrate, with or without pre-patterned Au electrodes. To maintain the integrity of the samples, these freshly prepared Hall devices, LTEM, and MFM specimens were promptly placed into a plastic box within the argon-filled glovebox and securely sealed with parafilm before removal. During subsequent measurements outside the glovebox, the exposure of the samples to air was minimized, and all the sample transfers were conducted within a confined timeframe of no more than 10 minutes. LTEM measurements were performed on Thermo Fisher Talos F200i at an acceleration voltage of 200 kV. The objective lens was used to apply magnetic field perpendicular to the sample plane by controlling the excitation current. The specimen was in-suit warmed and cooled by a liquid nitrogen double-tilt sample holder. In order to perform field-cooling with accurate temperature control, the sample was initially cooled to +20 K above the desired temperature at a rate of -5 K/min, and then slowly cooled down to the desired temperature at a rate of -1 K/min. The in-plane magnetization distribution map was reconstructed from the under- and over- focused images using the transport-of-intensity equation (TIE) approach. MFM measurements were performed by scanning probe microscopy (MFP-3D, Asylum Research), which equipped a low-moment magnetic tip (PPP-LM-MFMR, Nanosensors) and VFM3 component (Asylum Research). To protect samples from air degradation, Hall and LTEM measurements were conducted under vacuum conditions, while MFM measurements were carried out in an environment continuously flushed with argon gas to ensure effective protection.

**In-situ optical LTEM experiments.**

To directly visualize the fs laser writing of RT skyrmions in the 2D van der Waals $Fe_{2.84\pm0.05}GaTe_2$ nanoflakes, we performed the in-situ optical LTEM experiments in our homemade 4D-electron microscopy (Thermo Fisher Talos F200i), which enables single-shot fs laser pulse excitation on the sample under the Lorentz phase imaging mode. The Lorentz phase images were acquired under the Fresnel mode, in which the external perpendicular magnetic field was applied by the objective lens with controlled lens current. The fs laser system was triggered externally with a digital delay generator which outputted single-shot fs laser pulses with 520 nm wavelength, 300 fs pulse duration and 11 mJ/cm$^2$ fluence, where the laser spot size was adjusted to be 50 μm to ensure homogeneous illumination on the sample.

**Data availability**

The single crystal X-ray diffraction, Lorentz transmission electron microscopy and characterization data generated in this study are provided in the Supplementary Information file. The data that support the findings of this study are available from the corresponding authors upon reasonable request.

**Acknowledgement**

This work was supported by the National Key Research and Development Program of China at grant No. 2020YFA0309300, Science and Technology Projects in Guangzhou (grant No. 202201000008), the National Natural Science Foundation of China (NSFC) at grant No. 12304146, 11974191, 12127803, 52322108, 52271178, U22A20117 and 12241403, China Postdoctoral Science Foundation (2023M741828), Guangdong Basic and Applied Basic Research Foundation (grant No. 2021B1515120047 and 2023B1515020112), the Natural Science Foundation of Tianjin at grant No. 20JCJQJC00210, the 111 Project at grant No. B23045 , and the "Fundamental Research Funds for the Central Universities", Nankai University (grant No. 63213040, C029211101, C02922101, ZB22000104 and DK2300010207). This work was supported by the Synergetic Extreme Condition User Facility (SECUF).

**Author contributions**

X.W.F., Z.P.H. and Z.F.L. conceived the research project. Z.F.L., H.Z. and J.T.G. synthesized the crystals and characterized the structure. Z.F.L. and H.Z. performed the Lorentz phase microscopy, magnetic force microscopy and magneto-transport measurements. G.Q.L. performed the first-principles calculations. Z.F.L. performed the in-situ optical Lorentz phase microscopy. Z.F.L. and Q.P.W. did the micromagnetic simulation. The manuscript was drafted by Z.F.L., Z.P.H., Y.M.Z. and X.W.F. with contributions from H.Z., G.Q.L., J.T.G., Q.P.W., Y.D., Y.H., X.G.H., C.L., M.H.Q., X.S., R.C.Y., X.S.G., Z.M.L. and J.M.L. All the authors contributed to the discussion and revision of the manuscript.

**Competing interests**

The Authors declare no Competing Financial or Non-Financial Interests.

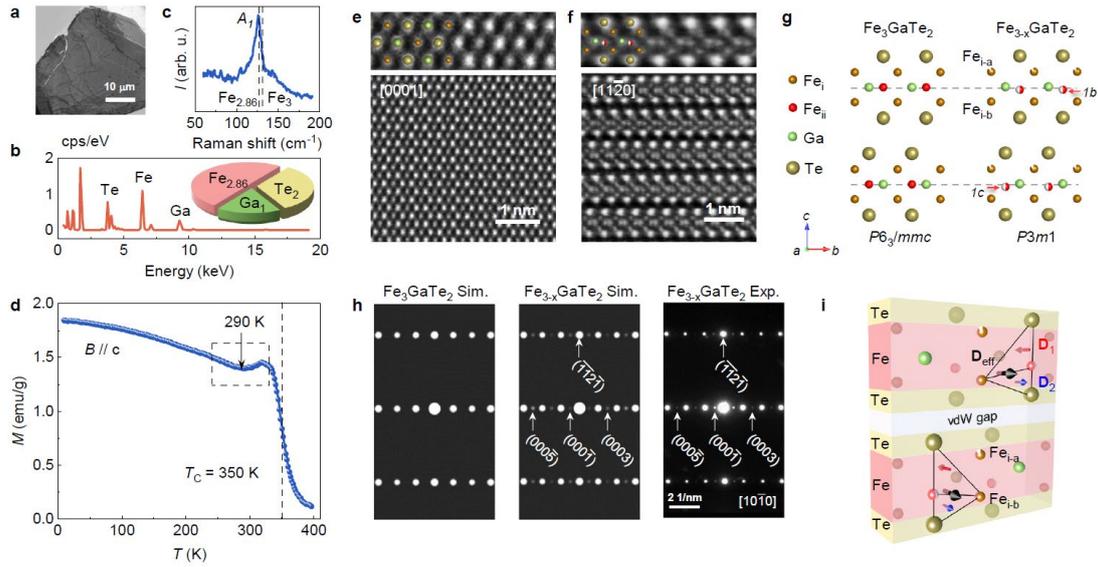

**Fig. 1 | Characterizations of Fe$_{2.84\pm0.05}$GaTe$_2$ single crystal. a** TEM image showing an exfoliated Fe$_{2.84\pm0.05}$GaTe$_2$ nanoflake on the Si$_3$N$_4$ membrane. **b** A typical EDX spectrum showing the chemical composition of the sample being Fe deficient. **c** Raman spectra for $A_1$ peak in Fe$_{2.84\pm0.05}$GaTe$_2$. The dashed line compares the $A_1$ peak shift with that of Fe$_3$GaTe$_2$[35]. **d** Temperature dependent magnetization curve $M(T)$ measured with field-cooled protocol at 10 mT. The dashed line marks the Curie temperature ($T_c$ = 350 K). The magnetization kink highlighted using dashed box indicates a rotation of magnetic easy axis at around 290 K. **e** HAADF images viewed along the [0001] zone axis. The enlarged panel shows hexagonal arrangement of Fe$_i$ (dark brown), Te (light brown), Ga (pale green) without noticeable lattice distortion. **f** HAADF images along the [11$\bar{2}$0] zone axis. The enlarged panel shows an obvious displacement of Fe$_{ii}$ (red) atoms from Ga-Ga plane. **g** The comparison of crystal structure between Fe$_3$GaTe$_2$ and Fe$_{2.84\pm0.05}$GaTe$_2$ from side view. The red arrows at right panel indicate the Wyckoff site of deviated Fe$_{ii}$ atoms at $1c$ and $1b$. **h** The comparison of simulated and experimental selected area electron diffractions (SAED) along [10$\bar{1}$0] zone axis. **i** Schematic illustration of DMI in asymmetric layers. The red arrow **D**$_1$ represents the direction of DMI vector in the upper triangle composed of Fe$_i$-Fe$_{ii}$-Te, while the blue arrow **D**$_2$ represents the lower part in the opposite direction. The black arrow **D**$_{eff}$ represents the sum of the non-zero DMI vector.

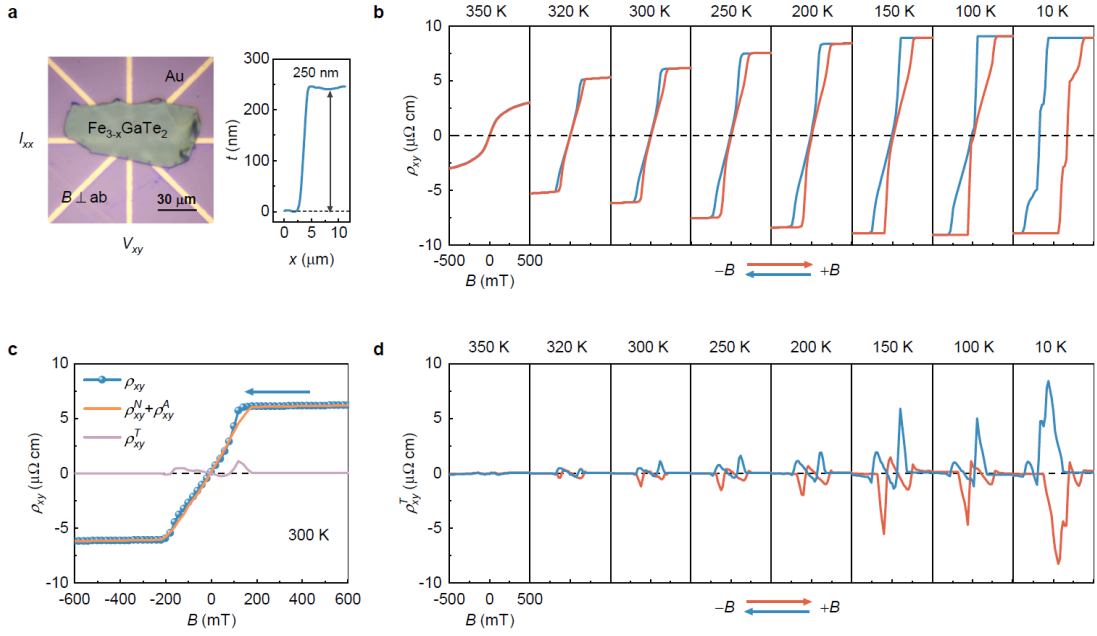

**Fig. 2 | Transport properties of Fe$_{2.84\pm0.05}$GaTe$_2$ nanoflake. a** Optical image of Fe$_{2.84\pm0.05}$GaTe$_2$ Hall device with a sample thickness of 250 nm. A current $I_{xx}$ was applied across the sample plane and transverse voltages $V_{xy}$ were measured simultaneously. **b** Magnetic hysteresis of Hall resistivity $\rho_{xy}$ at various temperatures from 350 to 10 K. Red (blue) curves were measured with increasing (decreasing) magnetic field. **c** Extraction procedure of normal Hall resistivity $\rho_{xy}^N$, anomalous Hall resistivity $\rho_{xy}^A$ and topological Hall resistivity $\rho_{xy}^T$ at 300 K. The blue arrow indicates the sweep direction of the magnetic field pointing downward. **d** Magnetic hysteresis of topological Hall resistivity $\rho_{xy}^T$ as a function of magnetic field at various temperatures from 350 K to 10 K. Red (blue) curves were extracted with increasing (decreasing) magnetic field.

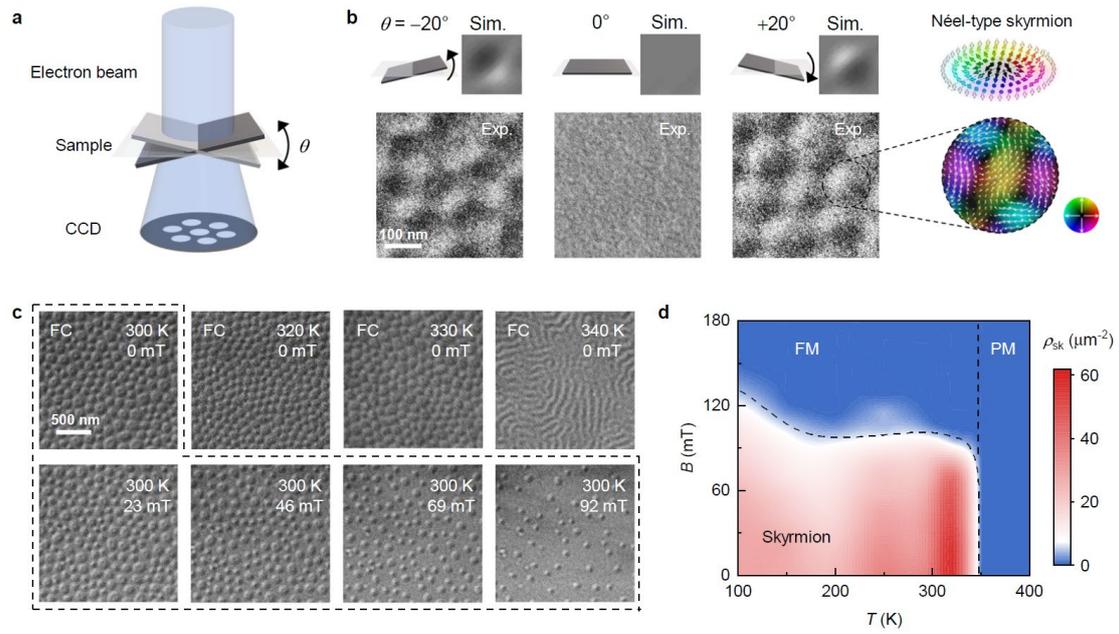

**Fig. 3 | LTEM measurements of Néel-type skyrmions in $Fe_{2.84\pm0.05}GaTe_2$. a** Schematic diagram of LTEM indicating the tilt angle $\theta$ with respect to the sample plane. All images were acquired with a defocus value of $d = 2$ mm. **b** Experimental and simulated Lorentz phase images of RT Néel-type skyrmions at ±20° and 0° tilt. The sample was previously applied field cooling (FC) with $B = 30$ mT. The illustration at the top utmost right panel shows a typical spin configuration of Néel-type skyrmions, while the bottom utmost right panel shows the corresponding magnetic induction field map at +20° tilt. **c** Temperature and magnetic field dependence of the skyrmion state. Note that the top row of Lorentz phase images at various temperatures were acquired individually after FC. The Lorentz phase images bounded by the dashed line represent the RT skyrmions evolution with increasing magnetic field. **d** Skyrmion phase diagram. The color indicates the skyrmion density $\rho_{sk}$. The black dashed line shows the boundary between skyrmion and ferromagnetic phase.

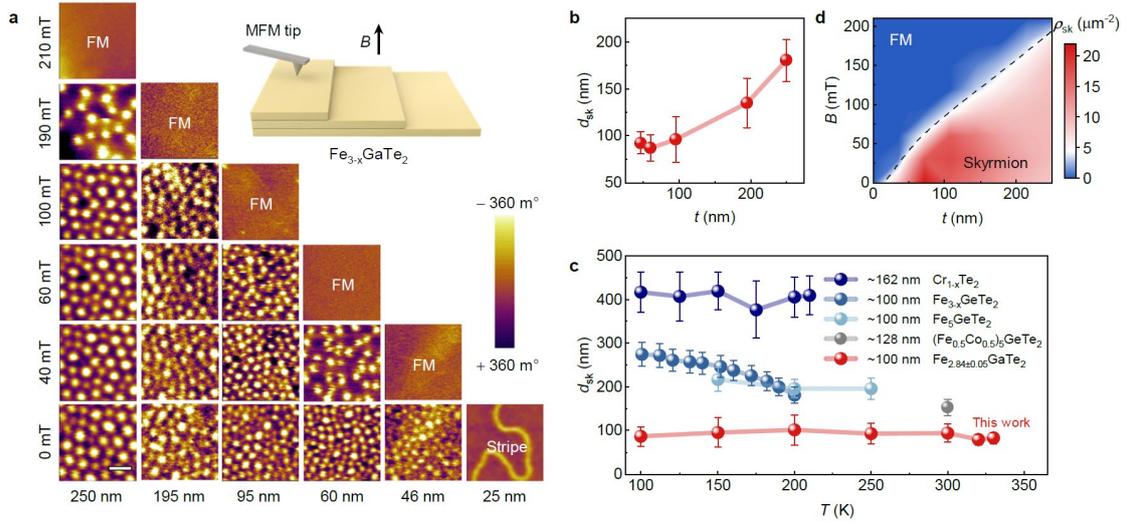

**Fig. 4 | Thickness-tunable skyrmions imaged by MFM. a** Typical MFM images of RT skyrmions taken at different thicknesses and external magnetic fields. In these images, bright contrast in the images corresponds to spin down, while dark contrast corresponds to spin up relative to the sample normal direction. The inset shows the schematic for the MFM experiment with magnetic field perpendicular to the sample plane. **b** RT skyrmion diameter $d_{sk}$ versus sample thickness $t$ at zero field. **c** Comparison for diameter $d_{sk}$ of field-free skyrmions and temperature $T$ for various 2D vdW materials with a thickness around 100 nm, including $Cr_{1+x}Te_2$[34], $Fe_{3-x}GeTe_2$[55], $Fe_5GeTe_2$[24], $(Fe_{0.5}Co_{0.5})_5GeTe_2$[31], and $Fe_{2.84\pm0.05}GaTe_2$ in this work. Error bars represent the standard error of skyrmion sizes averaged at various temperature. **d** Thickness and magnetic field dependence of RT skyrmion phase diagram. The color indicates the skyrmion density $\rho_{sk}$. The black dashed line shows the boundary between skyrmion and ferromagnetic phase.

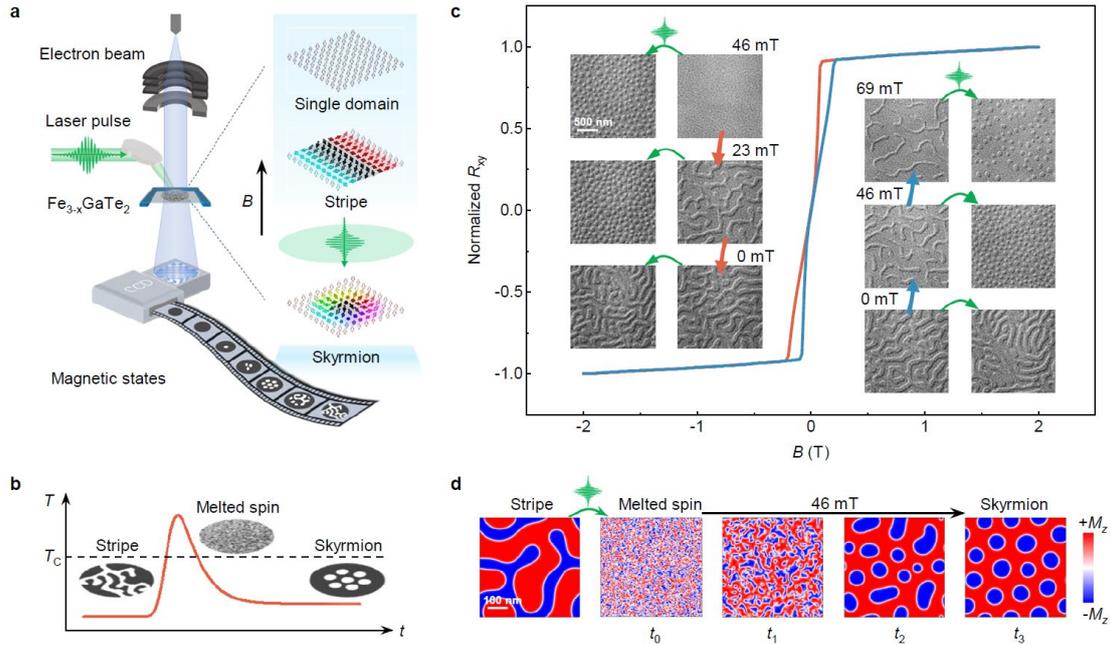

**Fig. 5 | Ultrafast fs laser writing of skyrmions in Fe$_{2.84\pm0.05}$GaTe$_2$. a** Schematic of the in-situ optical LTEM experiments. The sample was exposed to a single-shot fs laser pulse with an applied out-of-plane magnetic field. The defocused Lorentz phase images ($d = 2$ mm) captured the magnetic state at each step before and after the fs-laser pulse excitation. The right panel shows the typical spin configurations of stripe, single domain, and Néel-type skyrmions. **b** Schematic of the ultrafast demagnetization process for writing skyrmions from an initial stripe domain by a fs laser pulse. The sample is temporarily laser-heated above $T_c$ to melt the existing spin ordering, followed by quenching with external magnetic field to form a new spin ordering, skyrmions. **c** Illustration of single-shot laser pulse excitation from field-polarized magnetic state to subsequent laser-accessible magnetic states. The blue and red lines indicate the magnetization states during field swapping of Hall hysteresis curve. The inset shows corresponding Lorentz phase images before and after fs-laser pulses. **d** Simulated laser-induced stripe domain to skyrmion evolution process under an out-of-plane magnetic field of 46 mT. The magnetization along the $z$-axis ($M_z$) is represented by the color bar (+$M_z$ in red and −$M_z$ in blue).

Supplementary Information for

# Room-temperature sub-100 nm Néel-type skyrmions in non-stoichiometric van der Waals ferromagnet Fe$_{3-x}$GaTe$_2$ with ultrafast laser writability

**Authors:** Zefang Li[1#], Huai Zhang[2#], Guanqi Li[3#], Jiangteng Guo[1], Qingping Wang[4], Ying Deng[1], Yue Hu[1], Xuange Hu[1], Can Liu[1], Minghui Qin[2], Xi Shen[5], Richeng Yu[5], Xingsen Gao[2], Zhimin Liao[6], Junming Liu[2,7], Zhipeng Hou[2*], Yimei Zhu[8*] & Xuewen Fu[1,9*]

**Affiliations:**

[1] Ultrafast Electron Microscopy Laboratory, The MOE Key Laboratory of Weak-Light Nonlinear Photonics, School of Physics, Nankai University, Tianjin 300071, China

[2] Guangdong Provincial Key Laboratory of Optical Information Materials and Technology, Institute for Advanced Materials, South China Academy of Advanced Optoelectronics, South China Normal University, Guangzhou, 510006, China

[3] School of Integrated Circuits, Guangdong University of Technology, Guangzhou 510006, China

[4] College of Electronic information and automation, Aba Teachers University, Pixian Street, Wenchuan, 623002 China

[5] Beijing National Laboratory for Condensed Matter Physics, Institute of Physics, Chinese Academy of Sciences, Beijing, 100190 China

[6] State Key Laboratory for Mesoscopic Physics and Frontiers Science Center for Nano-optoelectronics, School of Physics, Peking University, Beijing 100871, China

[7] Laboratory of Solid State Microstructures and Innovation Center of Advanced Microstructures, Nanjing University, Nanjing 211102, China

[8] Condensed Matter Physics and Materials Science Department, Brookhaven National Laboratory, Upton, New York 11973, United States

[9] School of Materials Science and Engineering, Smart Sensing Interdisciplinary Science Center, Nankai University, Tianjin 300350, China

#These authors contribute equally to this work.

*Corresponding authors:
houzp@m.scnu.edu.cn (Z.H.); zhu@bnl.gov (Y.Z.); xwfu@nankai.edu.cn (X.F.).

## 1. Single crystal growth by Te-flux method.

**Supplementary Note 1:** We systematically grew a series of $Fe_{3-x}GaTe_2$ single crystals by varying the Fe content in the raw material composition, utilizing a Te-flux method. Subsequently, comprehensive energy-dispersive X-ray spectroscopy (EDS) mapping was conducted on the cleaved surfaces of these crystals to determine their chemical composition. To ensure the reliability of the EDS results, mapping was carried out at four distinct areas for each sample. Utilizing the Ga ratio as the normalization factor and ignoring the variation of Te content, their corresponding chemical composition were established. Table S1 provides a comprehensive overview of the raw material composition and the final crystal composition. It is clearly observed that when the raw Fe ratios fall below 0.8, the $Fe_{3-x}GaTe_2$ phase cannot be formed. Instead, a mixture of phases, including GaTe and $Ga_2Te_3$, is produced. When the raw Fe ratios are equal to or greater than 0.9, $Fe_{3-x}GaTe_2$ single crystals can be crystallized, with the Fe content in these single crystals increasing proportionally with the raw Fe ratios. However, an increase in the raw Fe ratios to 1.3 results in the formation of FeTe phases. In summary, we find that the chemical formulas for the crystals with the maximum and minimum Fe content correspond to $Fe_{2.84\pm0.05}GaTe_2$ and $Fe_{2.96\pm0.02}GaTe_2$, respectively. This result indicates that Fe vacancies always exist in the single crystals synthesized using a Te-flux method.

**Table S1.** Summary of the raw material composition and the final product for the growth of $Fe_{3-x}GaTe_2$ samples using the self-flux method.

| Molar ratio of Fe:Ga:Te | Mass of Fe (g) | Mass of Ga (g) | Mass of Te (g) | Product |
|---|---|---|---|---|
| 0.6:1:2 | 0.9349 | 1.9452 | 7.1200 | GaTe, $Ga_2Te_3$ |
| 0.7:1:2 | 1.0739 | 1.9153 | 7.0107 | |
| 0.8:1:2 | 1.2088 | 1.8864 | 6.9048 | |
| 0.9:1:2 | 1.3397 | 1.8583 | 6.8020 | $Fe_{2.84\pm0.05}GaTe_2$ |
| 1.0:1:2 | 1.4667 | 1.8310 | 6.7023 | $Fe_{2.91\pm0.04}GaTe_2$ |
| 1.1:1:2 | 1.5900 | 1.8046 | 6.6054 | $Fe_{2.95\pm0.03}GaTe_2$ |
| 1.2:1:2 | 1.7099 | 1.7789 | 6.5113 | $Fe_{2.96\pm0.02}GaTe_2$ |
| 1.3:1:2 | 1.8263 | 1.7539 | 6.4198 | FeTe |

## 2. Detailed EDS mapping and HAADF image.

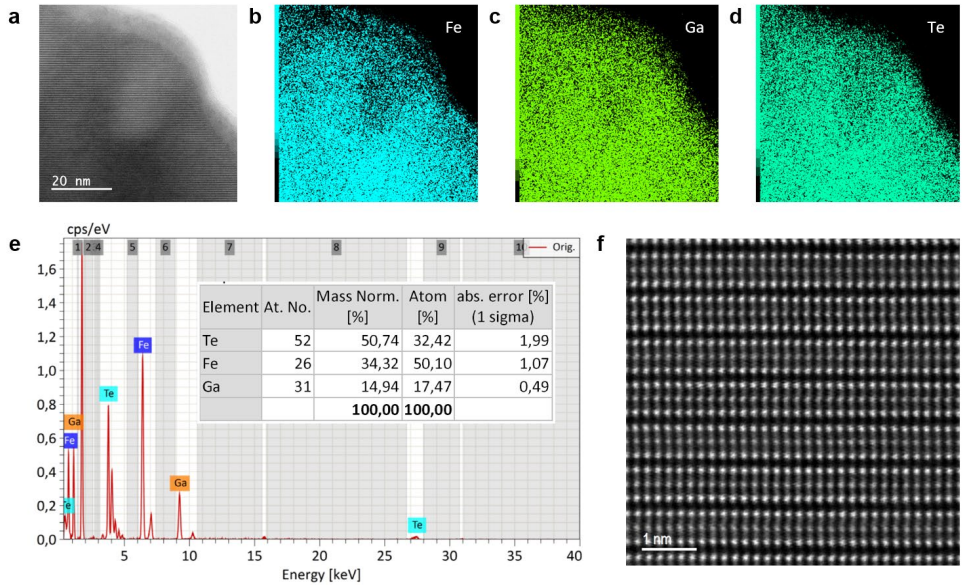

**Fig. S1. | EDS and HAADF-STEM characterizations.** **a** STEM images of a FIBed Fe$_{3-x}$GaTe$_2$ lamella along the [100] plane. **b-d** Corresponding EDS mapping of Fe, Ga, Te elements, respectively. **e** EDS spectrum and quantification results of Fe$_{2.84\pm0.05}$GaTe$_2$. **f** The HAADF image along [100] zone axis. The Fe-Ga-Te atom columns straightly lie along *c* axis, indicating no obvious lattice distortion in direction of *a* axis.

## 3. Detailed field-dependent magnetization.

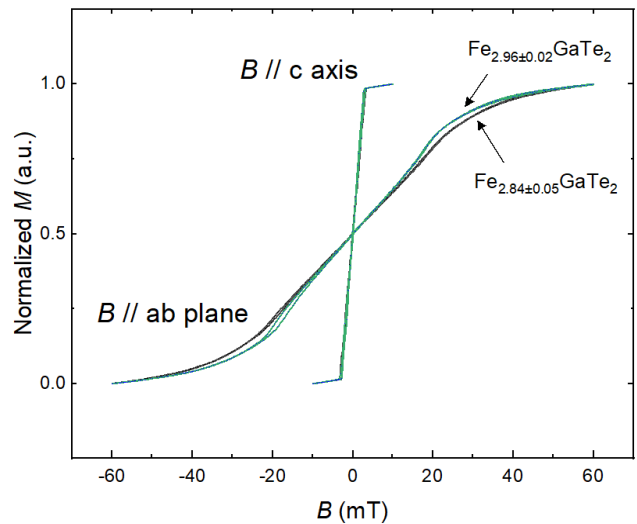

**Fig. S2. | Field-dependent magnetizations.** Room-temperature field-dependent magnetization curves for Fe$_{2.84\pm0.05}$GaTe$_2$ and Fe$_{2.96\pm0.02}$GaTe$_2$ samples.

## 4. Crystal structure of centrosymmetric Fe$_3$GaTe$_2$.

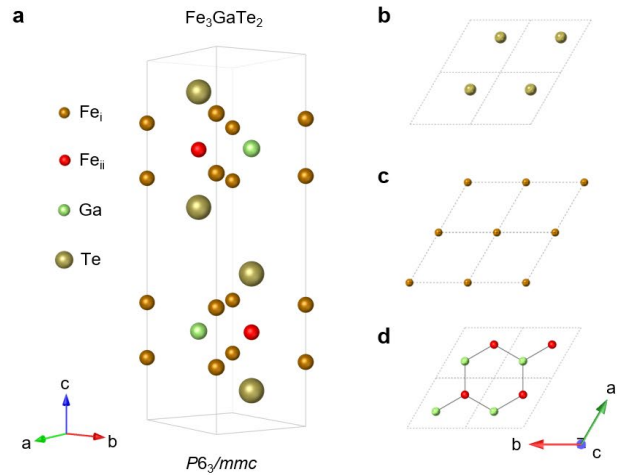

**Fig. S3. | Schematic diagram of Fe$_3$GaTe$_2$ crystal structure**. **a** Crystal structure of stoichiometric Fe$_3$GaTe$_2$. Each monolayer is sandwiched by Te-Fe$_i$-Fe$_{ii}$Ga-Fe$_i$-Te atom slices. **b** Top view of the Te atoms slice. **c** Top view of the Fe$_i$ atoms slice. **d** Top view of the Fe$_{ii}$Ga atoms slice.

## 5. STEM image analysis of Fe$_{2.84\pm0.05}$GaTe$_2$.

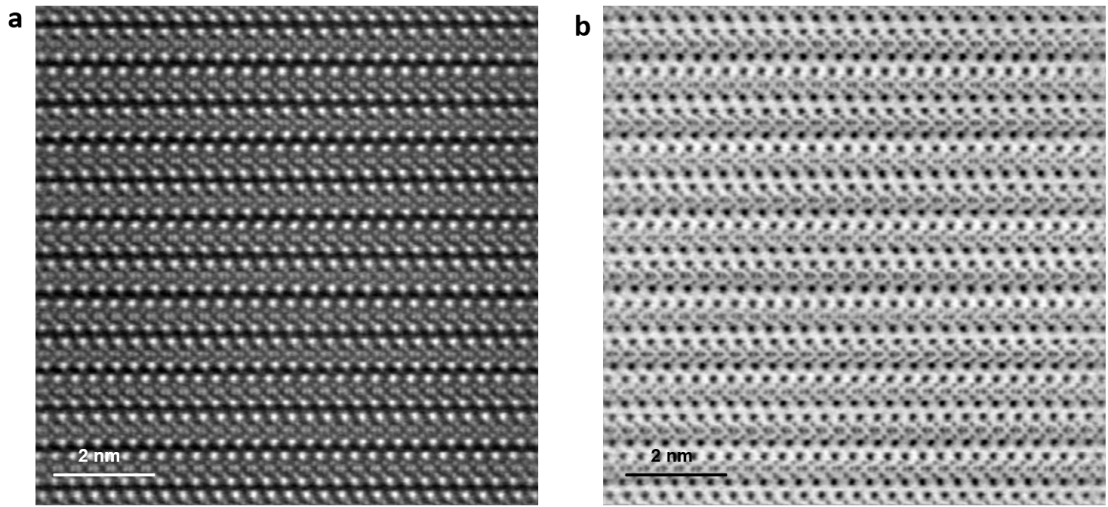

**Fig. S4. | HR-STEM characterizations**. **a** High-angle annular dark-field scanning transmission electron microscopy (HAADF-STEM) image and **b** annular bright-field scanning transmission electron microscopy (ABF-STEM) image of the Fe$_{2.84\pm0.05}$GaTe$_2$ sample along the $[11\bar{2}0]$ zone axis.

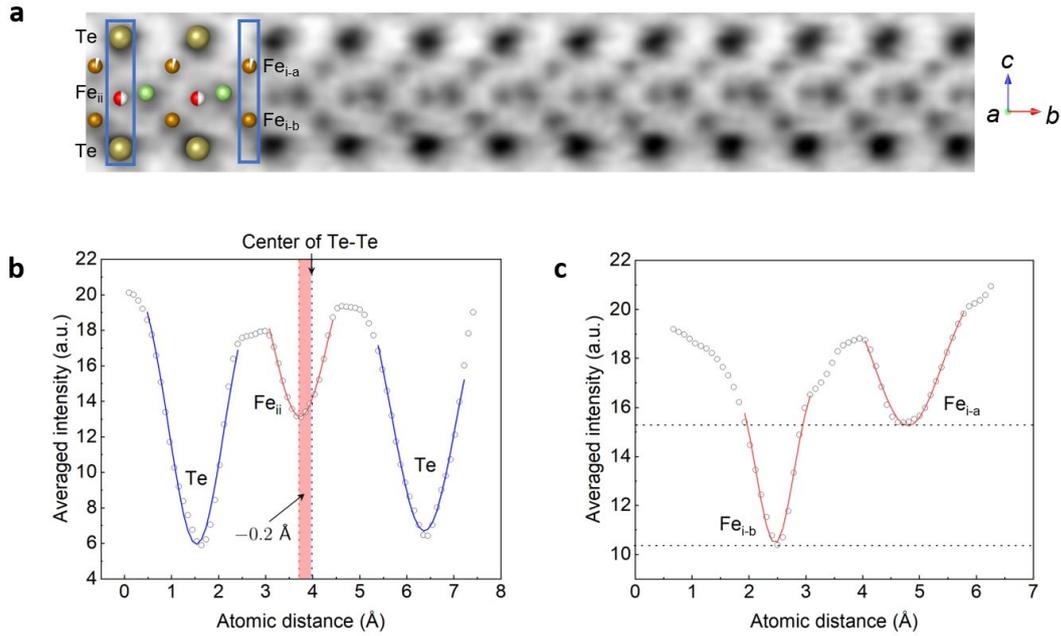

**Fig. S5. | Detailed STEM image analysis of Fe$_{2.84\pm0.05}$GaTe$_2$. a** Magnified ABF-STEM image of a single Fe$_{2.84\pm0.05}$GaTe$_2$ layer. **b** Integrated imaging intensity line profile along the *c*-axis within the area marked by the Te-Fe$_{ii}$-Te atoms in the left blue rectangle. The red region indicates the Fe$_{ii}$ deviation from the centers of Te-Te atoms. **c** Integrated imaging intensity line profile of Fe$_{i\text{-}a}$ and Fe$_{i\text{-}b}$ atoms along the c-axis within the area indicated by the right rectangle.

**Supplementary Note 2:** Fig. S4 shows HAADF- and ABF-STEM images of the Fe$_{2.84\pm0.05}$GaTe$_2$ sample along the [11$\bar{2}$0] zone axis, which provide clear view of the Fe$_i$ and Fe$_{ii}$ columns. To highlight the detailed information about the Fe$_{ii}$ columns, a magnified image was derived from the enclosed region of the ABF-STEM image, as shown in Fig. S5a. For a quantitative determination of the deviation of the Fe$_{ii}$ atoms, we focused on the region marked by left the blue rectangle in Fig. S5a comprising Te-Fe$_{ii}$-Te atoms. We then vertically integrated the corresponding imaging intensity line profile (Fig. S5b). By referencing the center of the two Te atoms, the deviation of the Fe$_{ii}$ atom towards the −$c$ direction was determined to be −0.20 Å. Utilizing the same procedure, we surveyed an area of 2 × 17 unit cells, yielding an average Fe$_{ii}$ deviation of −0.16 ± 0.06 Å.

Additionally, we observed that the image intensity of Fe$_{i-a}$ above Fe$_{ii}$ is weaker than that of Fe$_{i-b}$ below Fe$_{ii}$, as evident in the imaging intensity line profile of Fe$_{i-a}$ and Fe$_{i-b}$ atoms in Fig. S5c. Since imaging intensity is generally proportional to the number of projected atoms[1], the contrast difference between Fe$_{i-a}$ and Fe$_{i-b}$ indicates asymmetric site occupations, suggesting a small quantity of Fe vacancies in the Fe$_{i-a}$ site.

## 6. XRD refined crystal structure and simulated HR-STEM and SAED images.

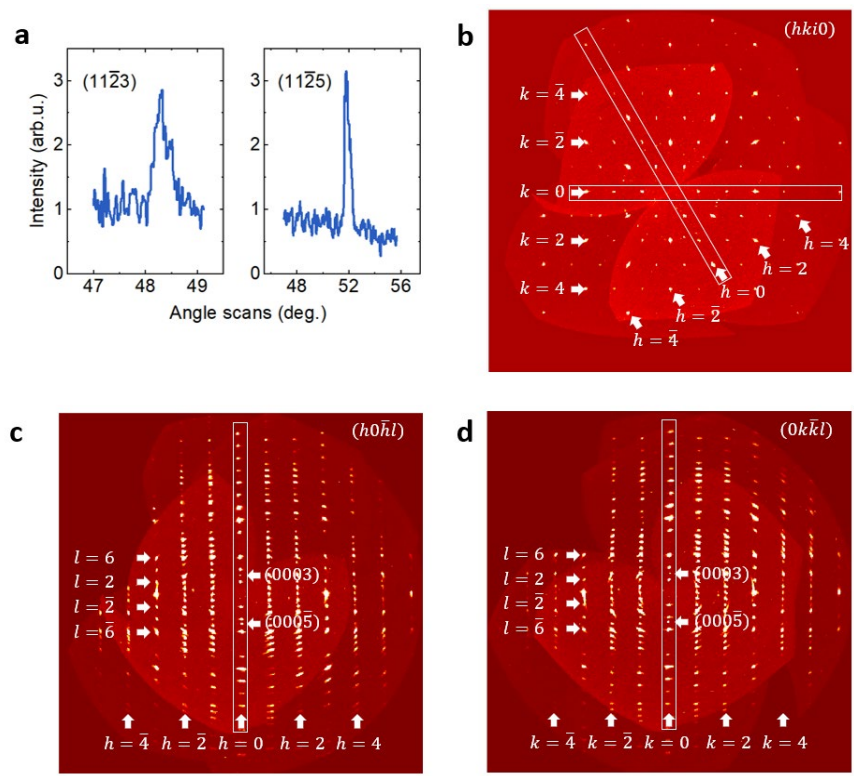

**Fig. S6. | Single-crystal XRD characterizations. a** XRD angle scans by four-circle diffractometer showing the typical reflections. Note that, the observed weak $(11\bar{2}3)$ and $(11\bar{2}5)$ reflections are forbidden for centrosymmetric space group $P6_3/mmc$. **b-d** The synthetic X-ray diffraction patterns of $Fe_{2.84\pm0.05}GaTe_2$ single crystal for **b** $(hki0)$, **c** $(h0\bar{h}l)$, and **d** $(0k\bar{k}l)$ reflections. Note that, the observed $(0003)$ and $(000\bar{5})$ reflections are forbidden for centrosymmetric space group $P6_3/mmc$.

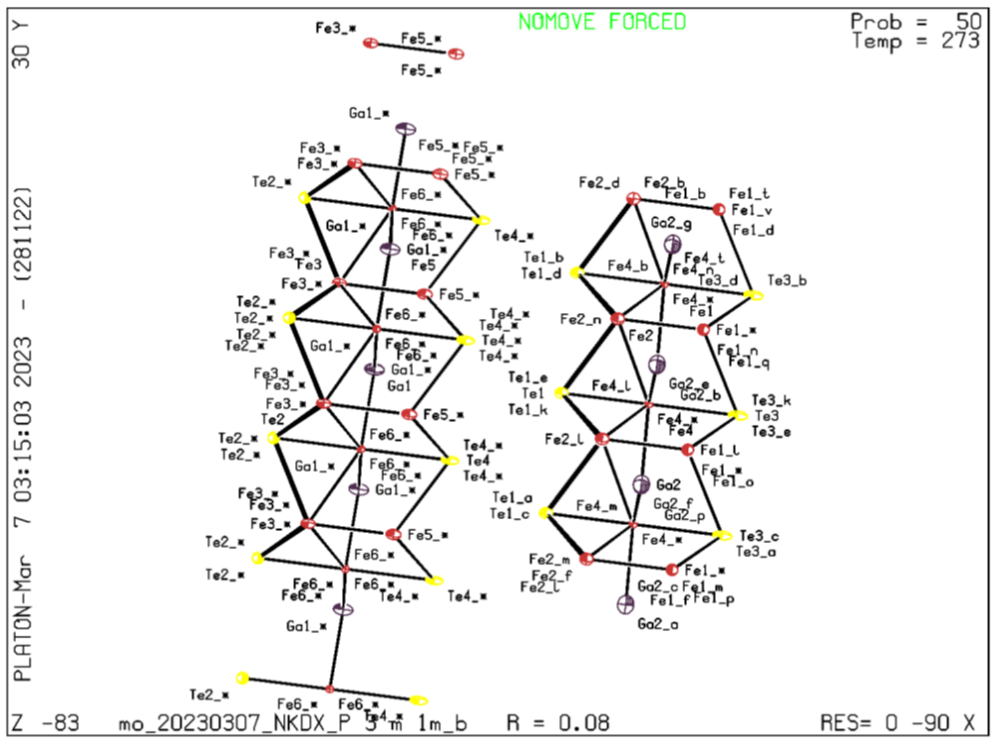

**Fig. S7. | XRD refined crystal structure of Fe$_{2.84\pm0.05}$GaTe$_2$.** The Fe$_{ii}$ sites (indicated as Fe6 and Fe4) deviate from the center position of Te-Te atoms, and both offsets in the top and bottom layers are the same direction, indicating a non-centrosymmetric crystal structure.

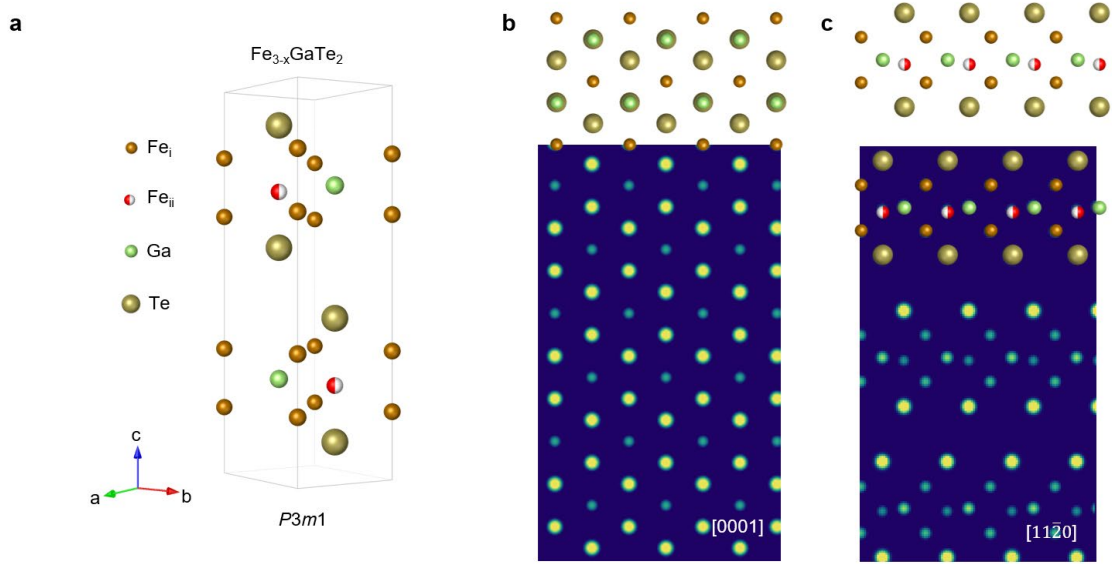

**Fig. S8. | Schematic diagram of Fe$_{2.84\pm0.05}$GaTe$_2$ crystal structure. a** Crystal structure of non-stoichiometric Fe$_{2.84\pm0.05}$GaTe$_2$ that is built with the refined Wyckoff site from XRD analysis. **b,c** Simulated HADDF images based on Dr. Probe [2], and the corresponding atom arrangements along the [0001] and [11$\bar{2}$0] zone axis.

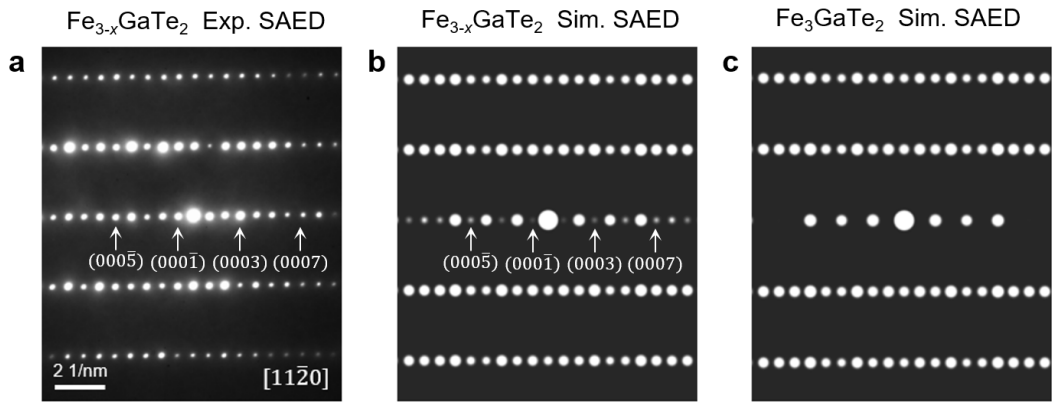

**Fig. S9. | Experimental and simulated $Fe_{3-x}GaTe_2$ SAED patterns. a** Selected Area Electron Diffraction (SAED) along the $[11\bar{2}0]$ zone axes. **b** Simulated SAED patterns based on the XRD refined non-centrosymmetric structure of $Fe_{3-x}GaTe_2$. **c** Simulated SAED patterns based on the prefect centrosymmetric structure of $Fe_3GaTe_2$.

**Supplementary Note 2**: A specimen of $Fe_{2.79}GaTe_2$ was used for the X-ray crystallographic analysis. The X-ray intensity data were measured ($\lambda = 0.71073$ Å). The frames were integrated with the Bruker SAINT software package using a narrow-frame algorithm. The integration of the data using a trigonal unit cell yielded a total of 1504 reflections to a maximum $\theta$ angle of 28.01° (0.76 Å resolution), of which 534 were independent (average redundancy 2.816, completeness = 99.3%, $R_{int}$ = 5.18%, $R_{sig}$ = 5.61%) and 455 (85.21%) were greater than $2\sigma(F^2)$. The final cell constants of $a$ = 4.0804(6) Å, $b$ = 4.0804(6) Å, $c$ = 16.127(4) Å, volume = 232.54(9) Å$^3$, are based upon the refinement of the XYZ-centroids of 2600 reflections above 20 $\sigma(I)$ with 5.052° < $2\theta$ < 56.00°. Data were corrected for absorption effects using the Multiscan method (SADABS). The ratio of minimum to maximum apparent transmission was 0.393.

The structure was solved and refined using the Bruker SHELXTL Software Package, using the space group $P\bar{3}m1$, with $Z = 2$ for the formula unit, $Fe_{2.79}GaTe_2$. The final anisotropic full-matrix least-squares refinement on $F^2$ with 38 variables converged at $R1$ = 8.33%, for the observed data and $wR2$ = 25.87% for all data. The goodness-of-fit was 1.192.

Table S2. Refined crystal data for $Fe_{3-x}GaTe_2$.

| Chemical formula | $Fe_{2.79}GaTe_2$ | |
|---|---|---|
| **Formula weight** | 480.80 g/mol | |
| **Temperature** | 273(2) K | |
| **Crystal system** | trigonal | |
| **Space group** | $P\bar{3}m1$ | |
| **Unit cell dimensions** | $a$ = 4.0804(6) Å | $\alpha$ = 90° |
| | $b$ = 4.0804(6) Å | $\beta$ = 90° |
| | $c$ = 16.127(4) Å | $\gamma$ = 120° |
| **Volume** | 232.54(9) Å$^3$ | |
| **Density (calculated)** | 6.867 g/cm$^3$ | |

## 7. Structural relaxation and electron density for Fe vacancies model.

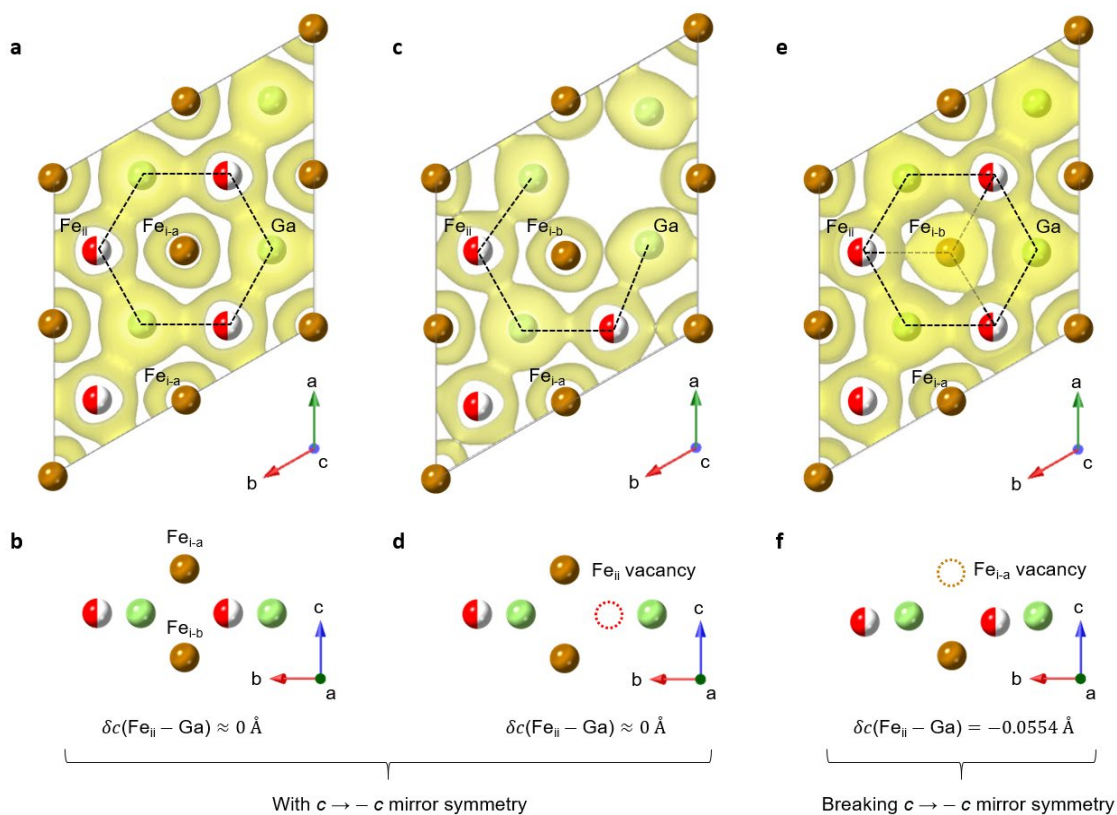

**Fig. S10. | First-principles calculations for Fe vacancies model.** The relaxed crystal structure and corresponding electron density of $Fe_{3-x}Ga$ atoms sliced from $Fe_{3-x}GaTe_2$ with **a**, **b** no vacancy, **c**, **d** $Fe_{ii}$ vacancy and **e**, **f** $Fe_{i-a}$ vacancy. The black dashed line indicates the chemical bonding with electron-density overlapping. The yellow-colored electron densities are shown at the same isosurface value.

**Supplementary Note 3**: Having established the crystal structure, we conducted first-principles calculations to analyze the impact of $Fe_{i-a}$ and $Fe_{ii}$ vacancies on $Fe_{ii}$ deviation. Starting with a perfect centrosymmetric $Fe_3GaTe_2$ lattice, we constructed a 2×2 supercell encompassing a total of four molecular layers. Within the bottommost layer, we systematically introduced three scenarios: no vacancy, $Fe_{i-a}$ vacancy, and $Fe_{ii}$ vacancy. Lattice relaxations were performed independently for the three distinct supercells. In order to illustrate the Fe vacancies and further compare the alterations in $Fe_{ii}$ chemical bonding, we presented the electron density of $Fe_{3-x}Ga$ atoms within the bottommost layer (Fig. S10).

Fig. S10a and S10b illustrate the scenario with no vacancy. The electron density

(colored in yellow) strongly overlaps between $Fe_{ii}$-Ga atoms, forming the $Fe_{ii}$-Ga honeycomb lattice plane with robust chemical bonding (highlighted by black dashed lines). Simultaneously, the $Fe_{i-a}$ and $Fe_{i-b}$ dimers are located at the center of the $Fe_{ii}$-Ga honeycomb lattice but do not bond with $Fe_{ii}$-Ga atoms. Consequently, the chemical bonding is mirror-symmetric along the $Fe_{ii}$-Ga plane, leading to the absence of $Fe_{ii}$ displacements. Thus, perfect $Fe_3GaTe_2$ exhibits a centrosymmetric crystal structure with $c \rightarrow -c$ mirror symmetry.

Fig. S10c and S10d illustrate the scenario with $Fe_{ii}$ vacancy. Although $Fe_{ii}$ vacancies cause deformation of the Ga electron density within the *ab* plane, there is still no overlapping with of $Fe_{i-a}$ and $Fe_{i-b}$ in the *c* direction. This observation indicates that the remaining $Fe_{ii}$ primarily forms bonds with Ga in the *ab* plane, with no displacement observed in the *c* direction.

Fig. S10e and S10f illustrate the scenario with $Fe_{i-a}$ vacancy. Apart from the strongly bonded $Fe_{ii}$-Ga honeycomb lattice, there is additional electron-density overlapping among the lower $Fe_{i-b}$ atom and its three nearest $Fe_{ii}$ atoms, while no overlapping between $Fe_{i-b}$ and its three nearest Ga atoms. Consequently, the newly formed chemical bonding (highlighted by black dashed line) will drag the $Fe_{ii}$ atoms displacing downwards to the lower $Fe_{i-b}$ atom. Notably, the determined deviation between nearest $Fe_{ii}$ and Ga atoms in the relaxed structure is −0.0554 Å. This value aligns with the analysis from ABF-STEM image, confirming the reliability of our DFT model. Based on the DFT results above, we conclude that the asymmetric vacancy of $Fe_{i-a}$ induces a displacement of $Fe_{ii}$ atoms towards the −*c* direction.

In addition, we calculated the vacancy formation energy for $Fe_i$ and $Fe_{ii}$. As for the calculation, we used a structure that repeats 3 times in the *c* direction and removed the bottommost $Fe_{i-a}$ or $Fe_{ii}$ atoms, and compared its total free energy with the parent structure. The calculated values of formation energy for $Fe_i$ and $Fe_{ii}$ was 2.96 eV/Fe and 2.86 eV/Fe, respectively. This result suggests that $Fe_{ii}$ vacancies are more likely to occur, aligning with experimental results.

Specific computational details: In all cases presented, the Vienna ab initio simulation package (VASP) was used with electron-core interactions described by the projector augmented wave method for the pseudopotentials, and the exchange correlation energy calculated with the generalized gradient approximation of the Perdew-Burke-Ernzerhof (PBE) form. The plane wave cutoff energy was 400 eV for all the calculations. In calculating the atomic shifts due to $Fe_{i-a}$ and $Fe_{ii}$ vacancies, we used a 2×2 supercell and removed the bottommost $Fe_{i-a}$ and $Fe_{ii}$ atoms. The Monckhorst-Pack scheme was used for the Γ-centred $12 \times 12 \times 1$ k-point sampling. All atoms' relaxations were performed until the force become smaller than 0.001 eV/A for determining the most stable geometries.

## 8. First-principles calculations for Dzyaloshinskii-Moriya interaction.

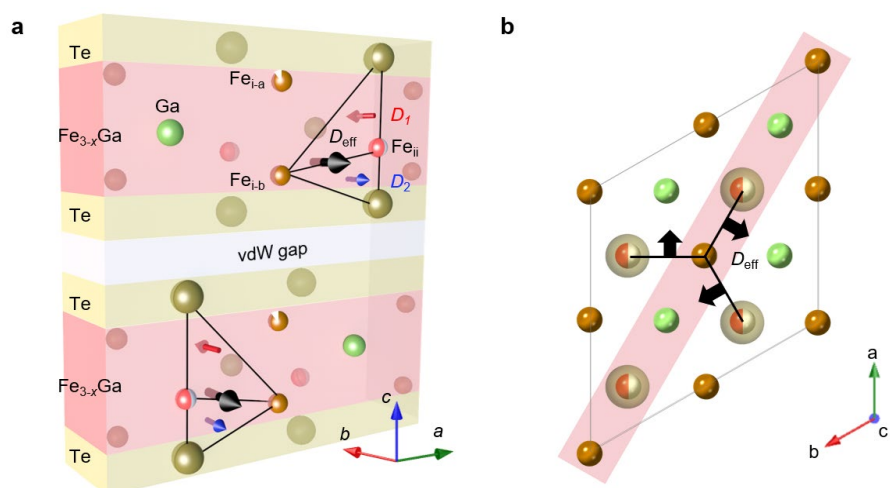

**Fig. S11. | Schematic illustration of DMI. a** Schematic illustration of DMI in asymmetric layers viewed from $[11\bar{2}0]$ zone axis. The red arrow $D_1$ represents the direction of DMI vector in the upper triangle composed of $Fe_i$-$Fe_{ii}$-Te, while the blue arrow $D_2$ represents the lower part in the opposite direction. The black arrow $D_{eff}$ represents the sum of the non-zero DMI vector. **b** Schematic illustration of DMI viewed from [0001] zone axis. The black arrow represents the effective net DMI vectors $D_{eff}$. And the red rectangular represents the slice of atoms in the left panel **a**.

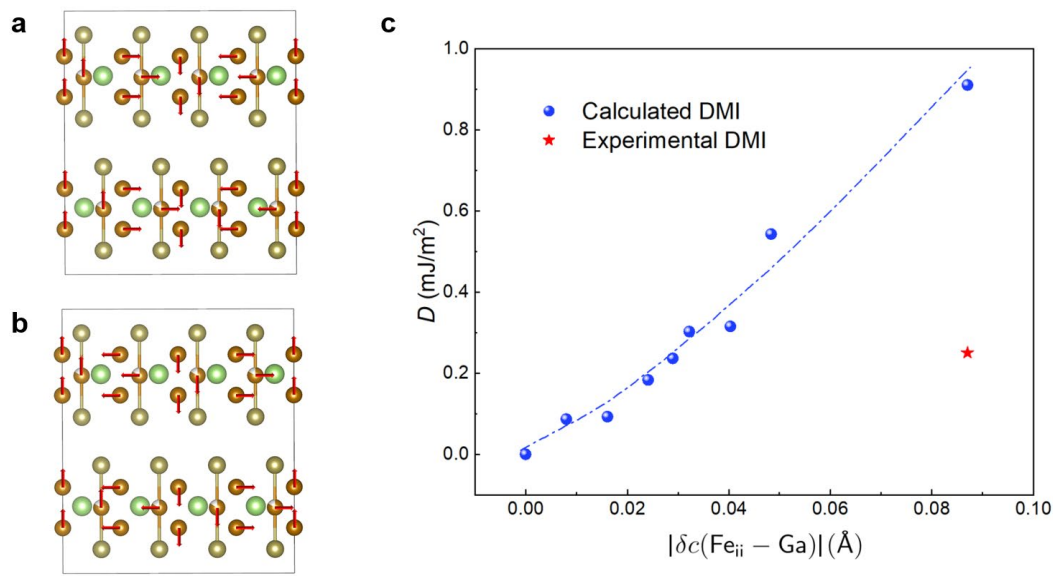

**Fig. S12. | First-principles calculations of DMI. a, b** Spin configurations implemented to calculate the DMI for clockwise (CW) and anticlockwise (ACW). **c** The calculated and experimental results for the relationship between the DMI and the $Fe_{ii}$ deviation.

**Supplementary Note 4**: In order to compare the DFT calculations with experimental observations, we investigated the relationship between $Fe_{ii}$ deviation value $\delta c(Fe_{ii} - Ga)$ and DMI constant $D$. As shown in Fig. S11 and S12, the crystal structures were built with different $Fe_{ii}$ deviation values $\delta c(Fe_{ii} - Ga)$, which are fixed during the two-step calculation for DMI values. The calculated $D$ with different $\delta c(Fe_{ii} - Ga)$ values are shown in Fig. S12c. On the one hand, the minimum $\delta c = 0$ indicate a centrosymmetric structure without $Fe_{ii}$ deviation. And the corresponding DMI value is $D = 0$ mJ/m$^2$, which is consistent with our expectation. On the other hand, the maximum $\delta c = -0.0871$ Å is equaling to the non-centrosymmetric structure determined by single-crystal XRD. And the corresponding DMI value is $D = 0.91$ mJ/m$^2$. Based on the measured magnetic domain width from LTEM experiments, the extracted $D = 0.25$ mJ/m$^2$ falls within the range of DFT calculated $D$ values.

Specific computational details: All calculations are based on the Vienna *ab initio* simulation package with electron-core interactions described by the projector augmented wave method for the Perdew-Burke-Ernzerhof (PBE) form[3-5], and the plane-wave cutoff energy is set to 400 eV. The 4 × 16 × 4 k-point grids were used to sample the Brillouin zones for the 4 × 1 × 2 Fe$_3$GaTe$_2$ supercell. To obtain the DMI vector, calculations were performed in two steps. First, structural relaxations were performed with Gaussian smearing until the forces become smaller than 0.001 eV/Å. Next, spin-orbit coupling was included in the calculation, and the total energy of the system was determined as a function of the spin configuration as shown in Fig. S12a, and $d_{\parallel}$ equals to $(E_{ACW} - E_{CW})/12$. The DMI constant $D$ was calculated using the equation $D = 3\sqrt{2}d/(N_F a^2)$, where $N_F$ is the number of atomic layers, $a$ is the lattice constant and $d_{\parallel}$ represents DMI strength[3]. In the second step, the EDIFF is set to 10$^{-8}$ eV and the tetrahedron method with Blöchl corrections was used to get an accurate total-energy. This method has been widely used in DMI calculations including bulk frustrated system and insulating chiral-lattice magnets[3-6].

## 9. Thickness dependent Hall resistivity at room-temperature.

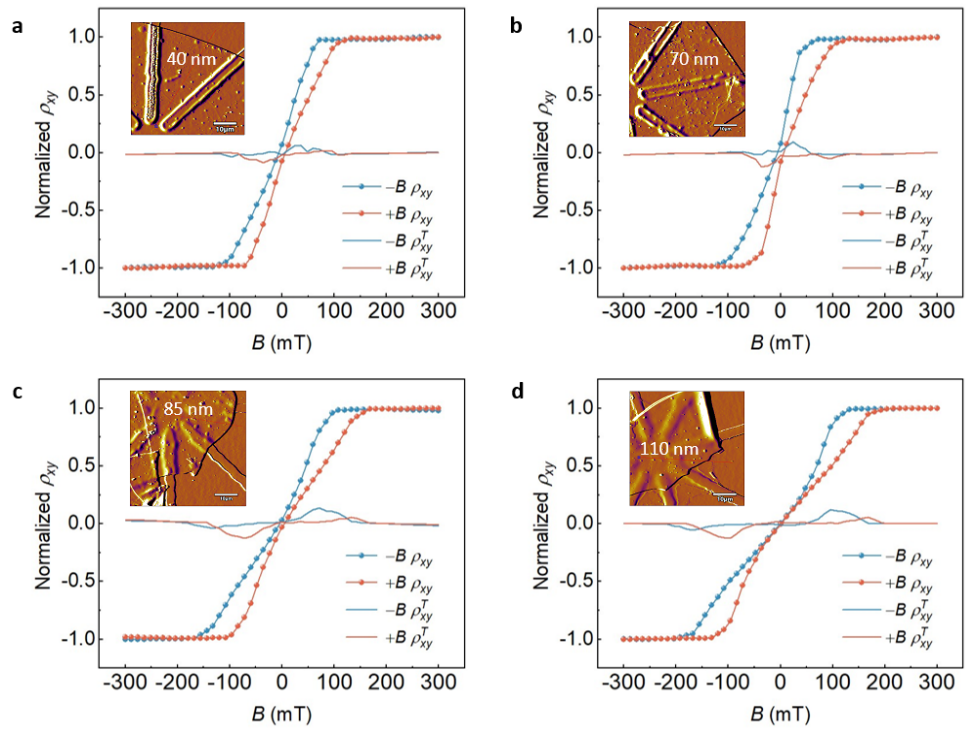

**Fig. S13. | Thickness dependent Hall measurements. a-d** Room-temperature magnetic hysteresis of Hall resistivity $\rho_{xy}$ and topological Hall resistivity $\rho_{xy}^T$ at various sample thickness from 40 nm to 110 nm. Red (blue) curves were measured with increasing (decreasing) magnetic field. The insets show the optical images of the Fe$_{2.84\pm0.05}$GaTe$_2$ Hall devices.

## 10. Detailed Lorentz phase images and AFM images.

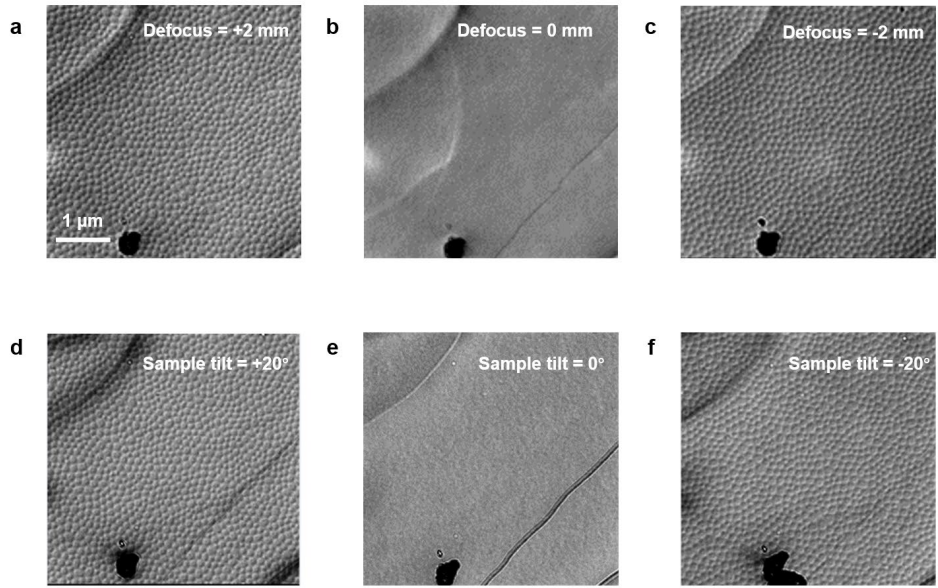

**Fig. S14. | Detailed LTEM characterizations. a** Over-focused, **b** in-focus, and **c** under-focused Lorentz phase images for room-temperature (RT) skyrmions after field-cooling procedure (30 mT). **d** +20°, **e** 0° and **f** −20° sample tilt of Lorentz phase images for RT Néel-type skyrmions. These images were taken at a defocus value of +2 mm.

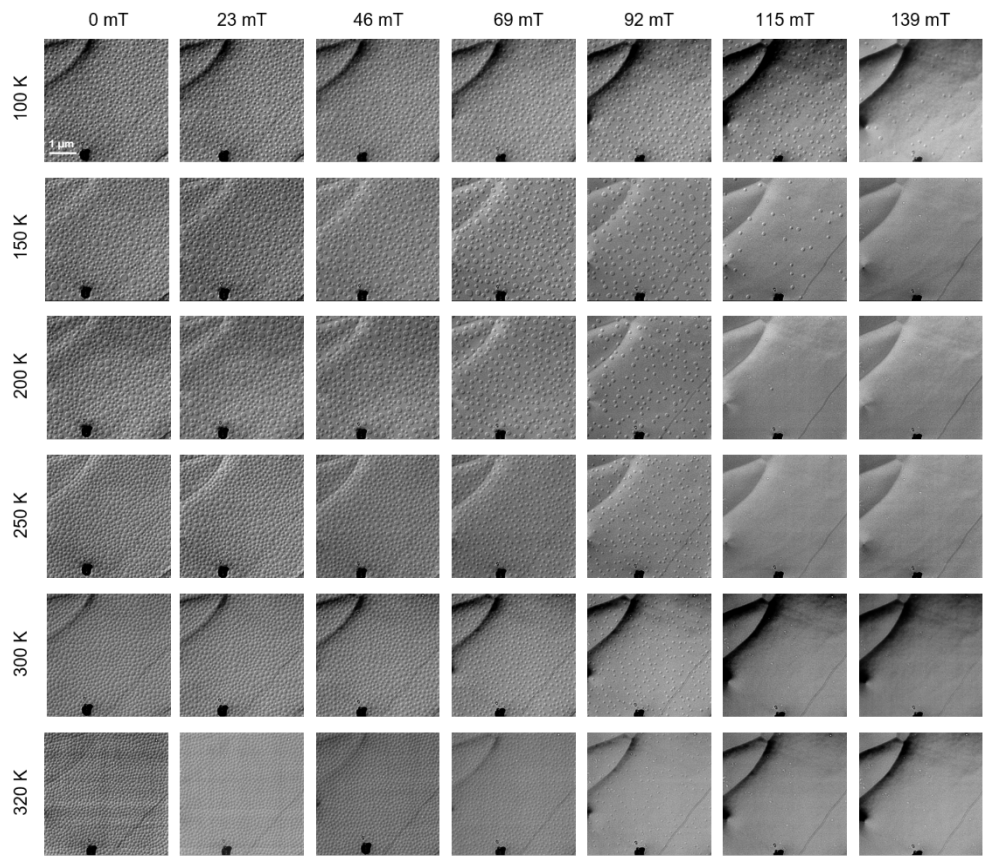

**Fig. S15. | Temperature and magnetic field dependent LTEM characterizations.** Series of Lorentz phase images taken at various magnetic fields and temperatures with a defocus value of +2 mm and +20° sample tilt.

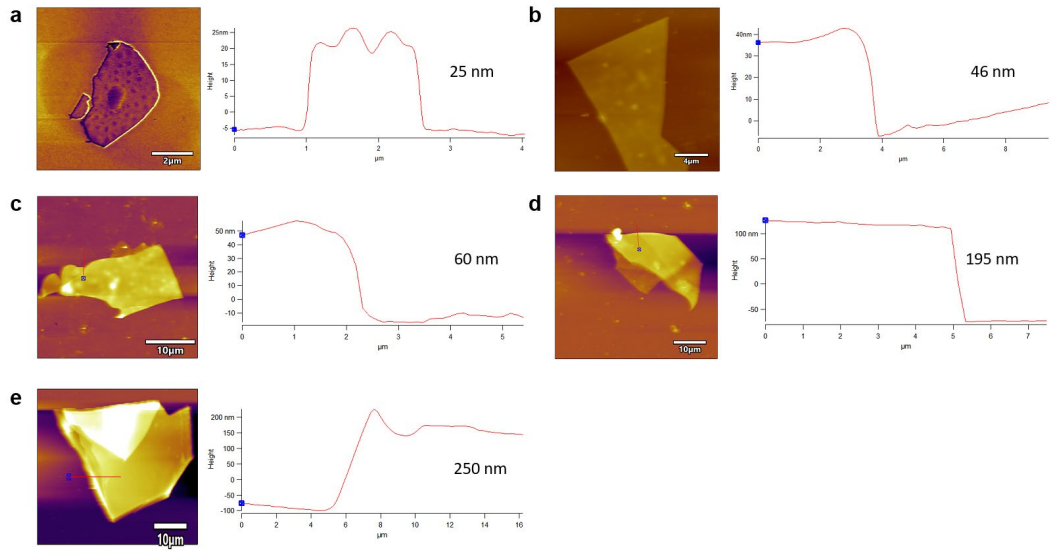

**Fig. S16. | AFM measurements. a-e** Sample morphology and thickness analysis by Atomic Force Microscope (AFM).

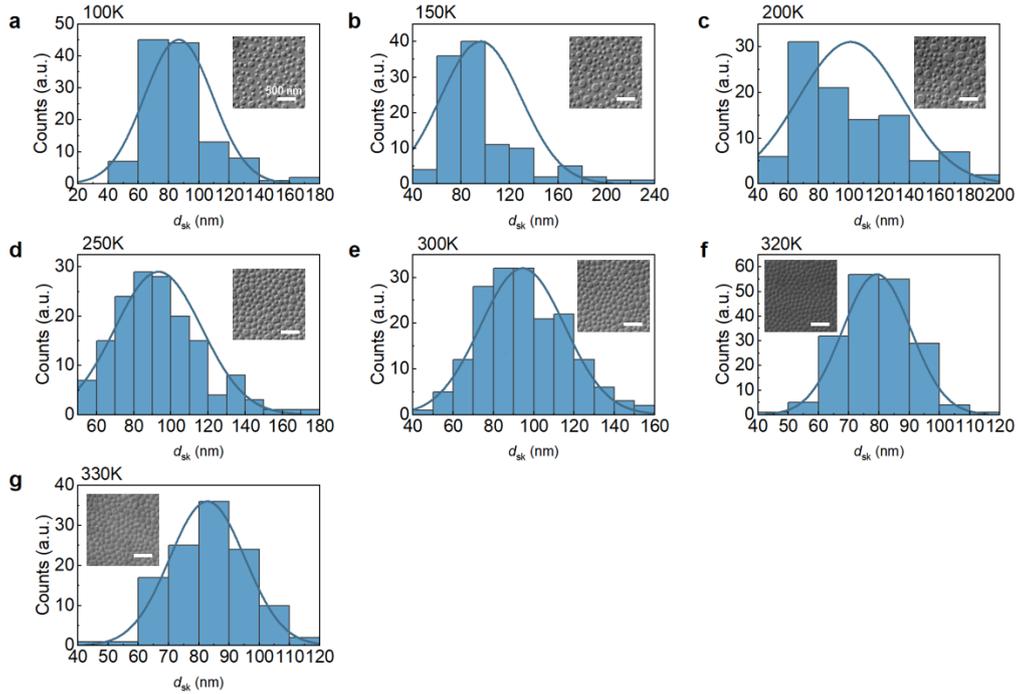

**Fig. S17. | Skyrmion size distribution. a-g** Statistics on the size distribution of field-free skyrmions at various temperatures.

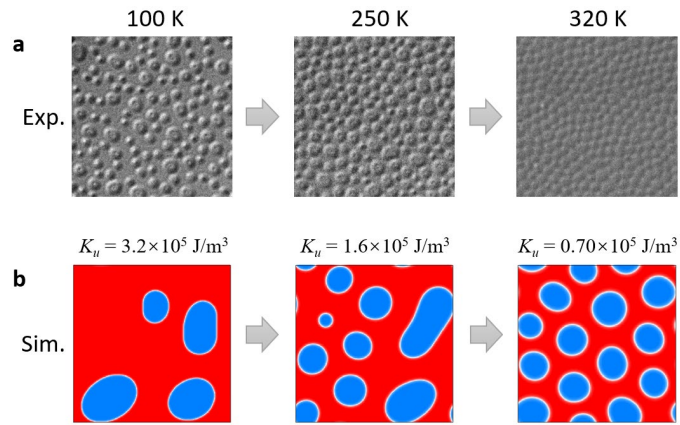

**Fig. S18. | Simulated skyrmion size distribution. a** Lorentz Phase images of zero-field skyrmions after 30 mT field cooling at 100 K, 250 K and 320 K. **b** Micromagnetic simulations of zero-field skyrmions after 30 mT field cooling with magnetic anisotropy constant $K_u = 3.2 \times 10^5$ J/m$^3$, $1.6 \times 10^5$ J/m$^3$ and $0.7 \times 10^5$ J/m$^3$.

**Supplementary Note 4**: In the LTEM experiments, we initially raised the sample temperature to above the Curie temperature, then cooled it to the target temperature with a 30 mT external magnetic field, and removed the external magnetic field to obtain the zero-field skyrmions. As shown in Fig. S17 and S18, the density of the zero-field skyrmions at low temperature (100 K) is relatively low, exhibiting non-uniformed size distribution. However, at higher temperatures of 250 K and 320 K, the zero-field skyrmion density gradually increases, and the size distribution becomes more uniform.

To clarify physical mechanism underlying the variation in skyrmion size at different temperatures, we conducted micromagnetic simulations of zero-field skyrmions after field cooling. As is known, the formation of skyrmions is determined by the magnetic parameters, such as magnetic anisotropy $K_u$, DMI constant $D$, saturation magnetization $M_s$, sample thickness $t$, and exchange stiffness $A$. However, increasing the sample temperature of $Fe_{3-x}GaTe_2$ results in the most notable variations in the reduction of magnetic anisotropy $K_u$. Therefore, as shown in Fig. S18, we simulated zero-field skyrmion after 30 mT field cooling with magnetic anisotropy constant $K_u = 3.2 \times 10^5$ J/m$^3$, $1.6 \times 10^5$ J/m$^3$ and $0.7 \times 10^5$ J/m$^3$, which is in accordance with the increasing of sample temperature. Our simulations demonstrate that with large $K_u$ at low temperature, the density of zero-field skyrmions is low. And the distant between the nearest skyrmions can be considerably large in certain regions, thus facilitating the expansion of skyrmion size upon the removal of the magnetic field. In contrast, with small $K_u$ at high temperature, the skyrmions exhibit a densely hexagonal arrangement to each other, which suppresses the extension of the skyrmions. Consequently, they remain uniformly distributed after removing the magnetic field.

Details about the simulation: To validate this experimental results, we conducted micromagnetic simulations with 30 mT field cooling process and then removing the external magnetic field. Default magnetic parameters used in the simulations include $A$ = 1.3 pJ/m, $K_u = 0.8 \times 10^5$ J/m$^3$, $M_s = 2.5 \times 10^5$ A/m, $D$ = 0.25 mJ/m$^2$, and slab geometries with dimensions of 512 × 512 × 64, with a mesh size of 2 × 2 × 2 nm. Periodic boundary conditions were taken into account for large-scale simulations.

## 11. Determination of magnetic parameters for $Fe_{2.84\pm0.05}GaTe_2$.

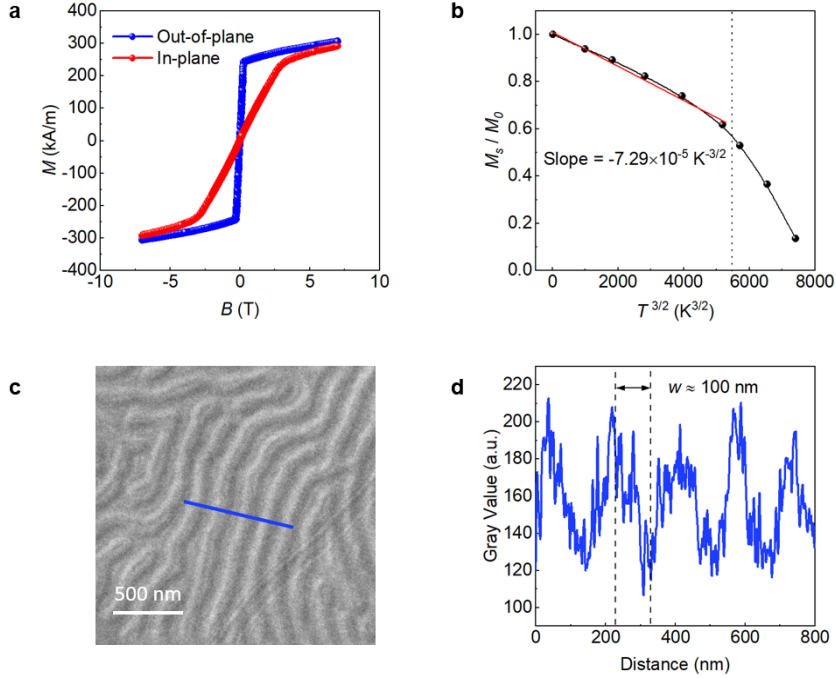

**Fig. S19. | Magnetic characterizations of $Fe_{2.84\pm0.05}GaTe_2$. a** Field dependence of magnetization at room temperature in bulk single crystal. **b** Normalized saturation magnetization $M_s/M_0$ versus $T^{3/2}$. The red line indicates the linear fitting. **c** Stripe domains at room temperature. **d** The corresponding gray value indicates domain wall period $w \approx 100$ nm.

**Supplementary Note 5**: Micromagnetic simulations were carried out using the GPU-accelerated micromagnetic simulation program Mumax$^3$. The slab geometries of dimensions were $512 \times 512 \times 64$ nm with a mesh size of $2 \times 2 \times 2$ nm. Periodic boundary conditions were taken into account for large-scale simulations. The total energy of the micromagnetic model contained exchange interactions, first-order uniaxial anisotropy, antisymmetric Dzyaloshinskii-Moriya interactions, magnetostatic energy and Zeeman energy.

In these simulations, the input parameters were determined from experimental measurements. As shown in Fig. S19a, the perpendicular magnetic anisotropy favors out-of-plane magnetization. We can thus estimate the saturation magnetization $M_s = 2.5 \times 10^5$ A/m. We consider the anisotropy constant in thin nanoflakes to be $K_u = 0.8 \times 10^5$ J/m$^3$.

The exchange stiffness *A* was determined by fitting the temperature dependence of magnetization with Bloch's $T^{3/2}$ law[7] (Fig. S19b):

$$M_s(T) = M_0 \left[1 - 0.0586 \frac{g\mu_B}{M_0} \left(\frac{kT}{d}\right)^{3/2}\right], \quad (1)$$

where $M_0$ is the saturation magnetization, *g* is the Landé *g*-factor, $\mu_B$ is the Bohr magneton, *k* is the Boltzmann constant, and *d* is the spin-wave stiffness that given by:

$$A = \frac{M_0}{2g\mu_B} d. \quad (2)$$

Therefore, we can extract the exchange stiffness A = 1.3 pJ/m.

The value of DMI constant *D* can be quantified by estimating the domain wall energy:

$$\sigma_{DW} = 4\sqrt{AK} - \pi|D|, \quad (3)$$

where $\sigma_{DW}$ is the domain wall surface energy density, which can be calculated by measuring the domain wall period *w* on the basis of a domain spacing model[8]:

$$\frac{\sigma_{DW}}{\mu_0 M_s^2 t} = \frac{w^2}{t^2} \sum_{odd\ n=1}^{\infty} \left(\frac{1}{(\pi n)^3}\right) [1 - (1 - 2\pi nt/w)\exp(-2\pi nt/w)], \quad (4)$$

where *t* is the sample thickness. As shown in Figs. S19c and S19d, the domain wall period *w* is approximately 100 nm. With a sample thickness *t* = 60 nm, we can calculate the DMI constant *D* = 0.25 mJ/m².

## 12. Micromagnetic simulations for skyrmion sizes.

Table S3. Magnetic parameters for skyrmion-host 2D materials.

| | $(Fe_{0.5}Co_{0.5})_5GeTe_2$ | $Fe_{3-x}GaTe_2$ |
|---|---|---|
| $D$ (mJ/m$^2$) | 0.90 | 0.25 |
| $M_s$ ($\times 10^5$ A/m) | 3.0 | 2.5 |
| $t$ (nm) | $\geq$110 | $\geq$46 |
| $K_u$ ($\times 10^5$ J/m$^3$) | 2.4 | 0.8 |
| $A$ (pJ/m) | 4.0 | 1.3 |
| $T$ (K) | 300 | 300 |

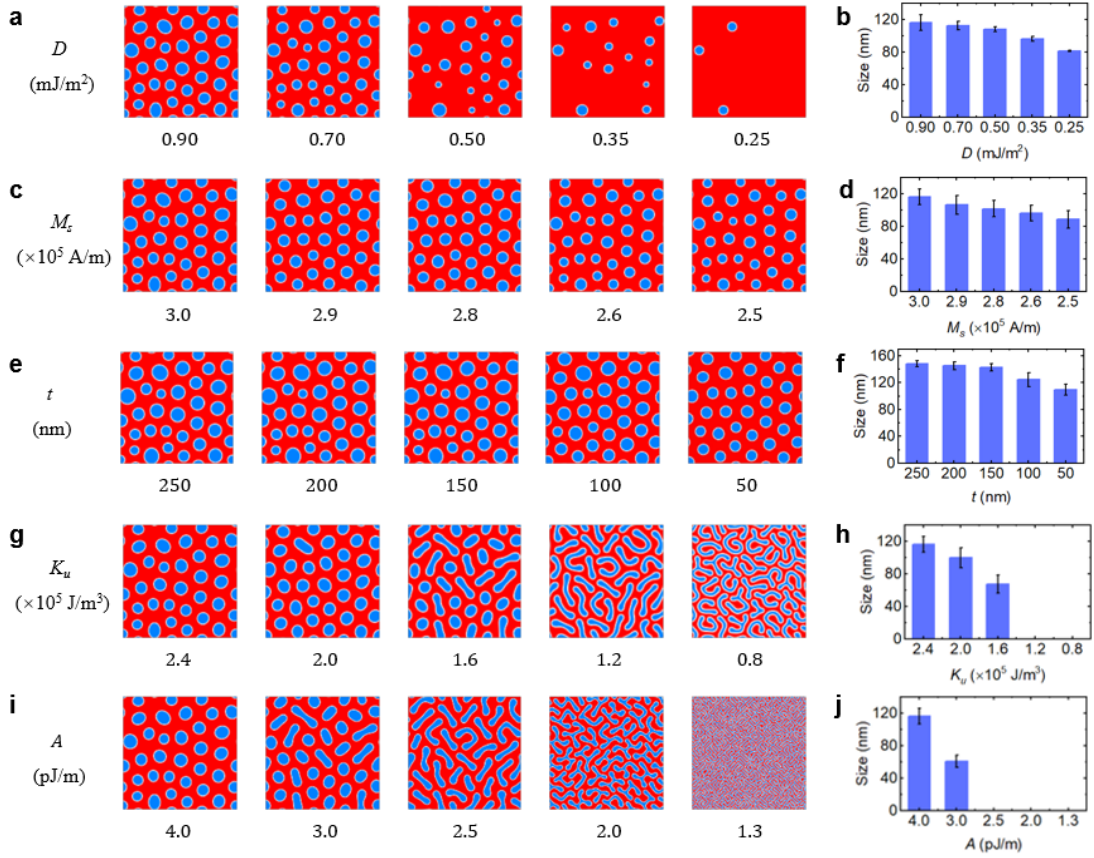

**Fig. S20. | Micromagnetic simulations for skyrmion sizes.** Variation of simulated magnetic structure and corresponding skyrmion sizes with varied DMI constant $D$ (**a, b**), saturation magnetization $M_s$ (**c, d**), sample thickness $t$ (**e, f**), magnetic anisotropy constant $K_u$ (**g, h**), and exchange stiffness $A$ (**i, j**).

**Supplementary Note 6**: It is widely recognized that the skyrmion size is greatly influenced by the interplay among various factors, including DMI, magnetic dipolar interaction, magnetic exchange interaction, saturation magnetization, and magnetic anisotropy. Utilizing the experimentally established magnetic parameters associated with these magnetic interactions in different vdW magnets (Table S3), we conducted micromagnetic simulations to clarify the contributions of these factors to the skyrmion size.

In our initial simulation, we modeled the zero-field skyrmion state following a 60 mT field cooling, adopting the magnetic parameters of $(Fe_{0.5}Co_{0.5})_5GeTe_2$ as a reference[9]. The simulation results, as depicted in Fig. S20, revealed a high-density of skyrmions with an average size of approximately 116 nm. Subsequently, by initiating the simulation with this skyrmion state and progressively reducing the DMI constant $D$ to match that of $Fe_{3-x}GaTe_2$ while keeping other magnetic parameters constant (Fig. S20a and S20b), we observed a significant decrease in the skyrmion size to 81 nm at $D$ = 0.25 mJ/m$^2$. Employing a similar approach, we further simulated the magnetic domain states by varying the saturation magnetization $M_s$, sample thickness $t$, magnetic anisotropy constant $K_u$ and exchange stiffness $A$ towards those of $Fe_{3-x}GaTe_2$. It is clearly demonstrated that each parameter reduction leads to a decrement in the skyrmion size. Thus, our simulations suggest that smaller magnetic parameters such as $D$, $M_s$, $t$, $K_u$, and $A$ in $Fe_{3-x}GaTe_2$ (in comparison to $(Fe_{0.5}Co_{0.5})_5GeTe_2$), contribute to the reduction in skyrmion size.

Details about the micromagnetic simulation: Micromagnetic simulations were carried out using the GPU-accelerated micromagnetic simulation program Mumax$^3$. Unless specified otherwise, default magnetic parameters used in the simulations include $A$ = 4.0 pJ/m, $K_u$ = 2.4 × 10$^5$ J/m$^3$, $M_s$ = 3.0 × 10$^5$ A/m, $D$ = 0.90 mJ/m$^2$, and slab geometries with dimensions of 512 × 512 × 64, with a mesh size of 2 × 2 × 2 nm. Periodic boundary conditions were taken into account for large-scale simulations. The initial skyrmion state was relaxed from a random state with 60 mT magnetic field. Employing the initial skyrmion state as the input, we systematically varied the magnetic

parameters—$D$, $M_s$, $t$, $K_u$, and $A$ individually—subsequently allowing the magnetization to evolve into a stabilized state through relaxation processes.

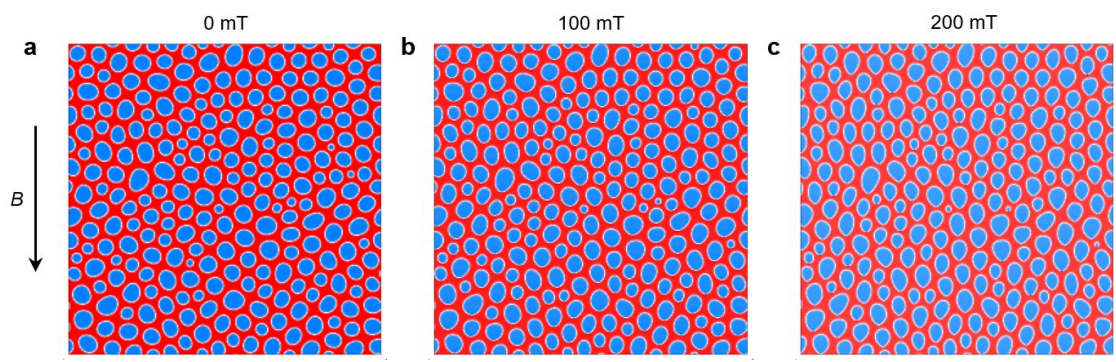

**Fig. S21. | Micromagnetic simulations for skyrmion shapes.** Micromagnetic simulations of skyrmion evaluation with in-plane magnetic field $B$ at **a** 0 mT, **b** 100 mT and **c** 200 mT, respectively. The shape of skyrmion transforms gradually from circular to elliptical.

**Supplementary Note 7**: As observed in our previous LTEM and MFM experiments (Fig. 3 and Fig. 4 in the main text), the out-of-plane magnetic field $B$ did not change the shape of skyrmions. The skyrmions remained circular and exhibited a reduction in size as increasing out-of-plane $B$. In order to further investigate the influence of skyrmion shape with in-plane magnetic field $B$, we performed micromagnetic simulations and the results are shown in Fig. S21. It can be seen that as the in-plane magnetic field increases, skyrmions transforms from circular to elliptical at $B = 200$ mT, with the elongated shape along the direction of the in-plane magnetic field.

Details about the simulation: Micromagnetic simulations were carried out using the GPU-accelerated micromagnetic simulation program Mumax$^3$. Default magnetic parameters used in the simulations include $A = 1.3$ pJ/m, $K_u = 0.8 \times 10^5$ J/m$^3$, $M_s = 2.5 \times 10^5$ A/m, $D = 0.25$ mJ/m$^2$, and slab geometries with dimensions of $1024 \times 1024 \times 16$, with a mesh size of $2 \times 2 \times 4$ nm. Periodic boundary conditions were taken into account for large-scale simulations. The initial zero-field skyrmion state was relaxed from a random state with 100 mT magnetic field. Employing the initial skyrmion state as the input, we systematically applied an in-plane magnetic field along $y$ axis, and relaxed the magnetization to a stable state.

## 4  Calculation of two temperature model.

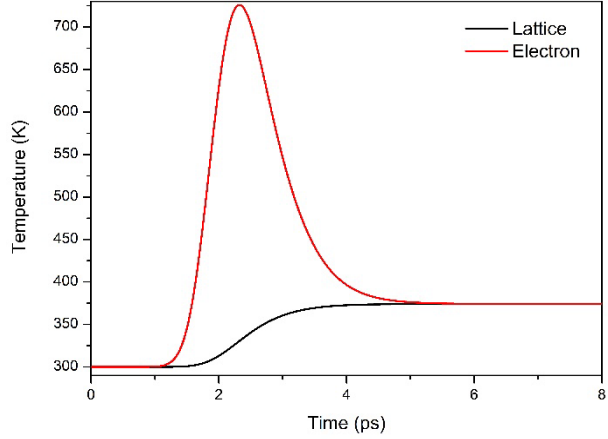

**Fig. S22. | Two-temperature model.** The simulated electron and phonon temperature versus delay times of $Fe_{2.84\pm0.05}GaTe_2$.

**Supplementary Note 8**: A phenomenological two-temperature model (2TM) model was utilized to describe the ultra-fast demagnetization dynamics during the fs-laser excitation on $Fe_{2.84\pm0.05}GaTe_2$. The 2TM includes the electron temperature $T_{elec}$ and phonon (lattice) temperature $T_{phon}$, which can be described by the following equations:

$$C_e \frac{\partial T_{elec}}{\partial t} = -G_{ep}(T_{elec} - T_{phon}) + S(t), \qquad (5)$$

$$C_p \frac{\partial T_{elec}}{\partial t} = -G_{ep}(T_{elec} - T_{phon}), \qquad (6)$$

where $C_e$ and $C_p$ donate the heart capacities of the electron and phonon systems that independent of the temperature, $G_{ep}$ is the electron-phonon coupling parameter, and $S(t)$ is the power of the single-shot laser pulse. The best fit variables of 2TM for $Fe_{2.84\pm0.05}GaTe_2$ are referred to the data of $Fe_3GeTe_2$ [10].

Based on Landau-Lifshitz-Gilbert equation, these initial field-polarized stripe domain state was relaxed by varying the out-of-plane magnetic field. As indicated by 2TM model (Fig. S22), the laser writing of skyrmion is dominated by thermal effect similar to that of field-cooling process. Therefore, we took a random spin configuration as the laser-melted spin state, and the subsequent skyrmion states after laser were relaxed with applying out-of-plane magnetic field. Furthermore, a damping constant $α$ = 0.5 was used to quickly reach the equilibrium magnetic state.

## 5 Detailed in-situ optical LTEM experiments.

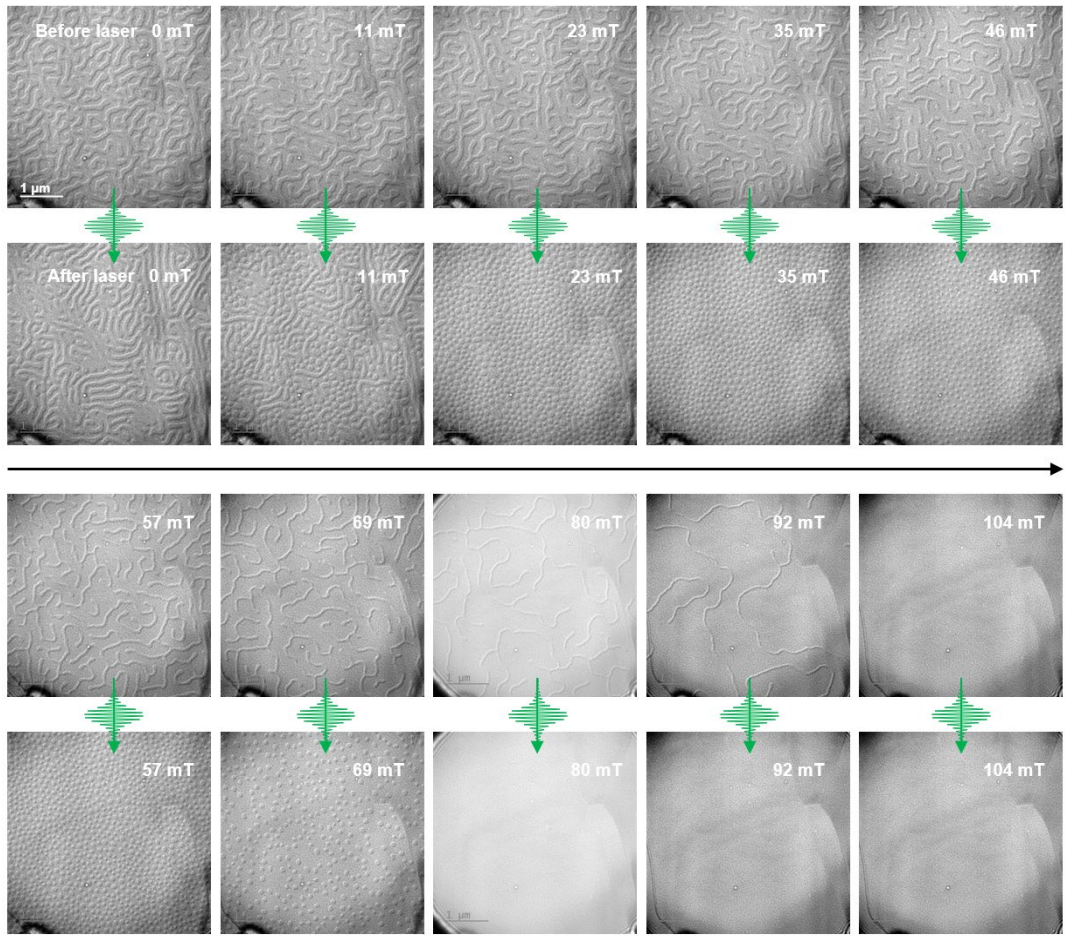

**Fig. S23. | In-situ laser writing of skyrmions with increasing *B*.** Step-by-step single-shot laser pulse excitation (11 mJ/cm$^2$) on a specific field-polarized magnetic state to determine the subsequent laser-accessible magnetic states. These images were recorded with increasing magnetic field.

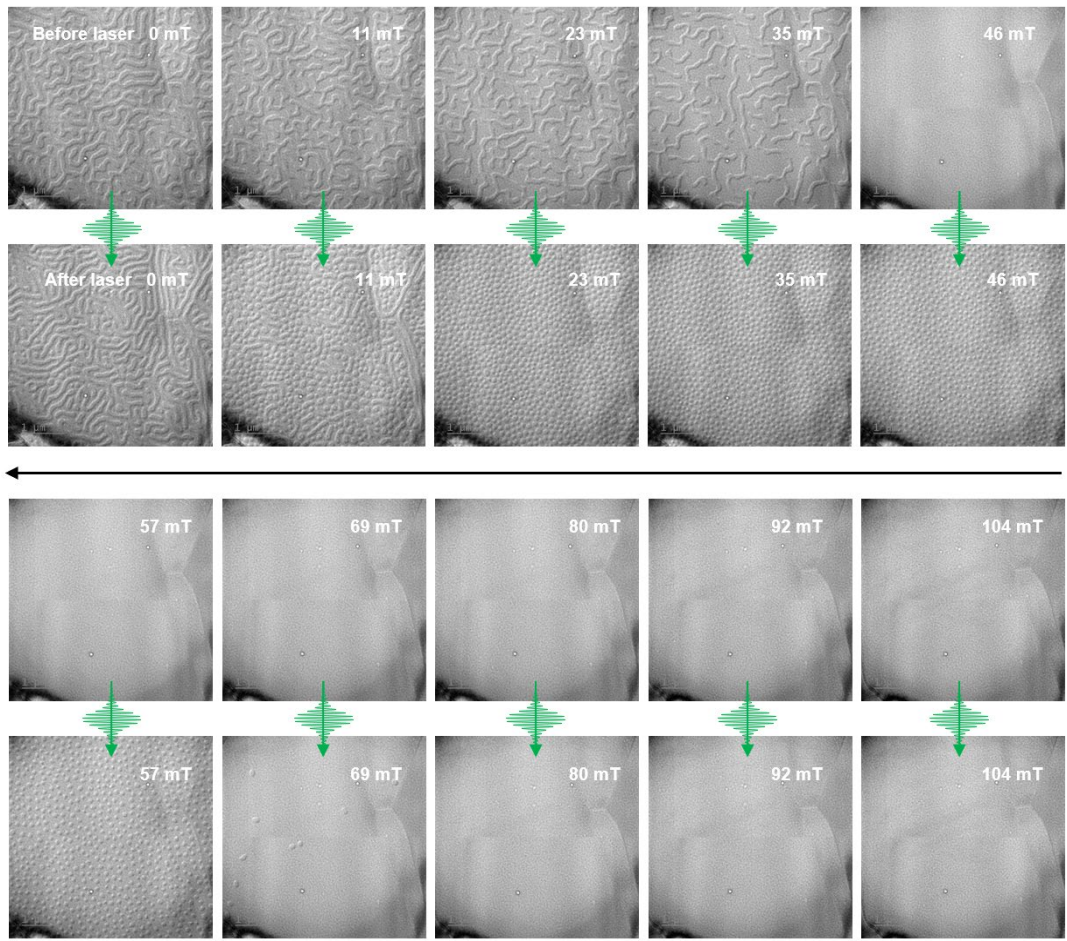

**Fig. S24. | In-situ laser writing of skyrmions with decreasing *B*.** Step-by-step single-shot laser pulse excitation (11 mJ/cm$^2$) on a specific field-polarized magnetic state to determine the subsequent laser-accessible magnetic states. These images were recorded with decreasing magnetic field.

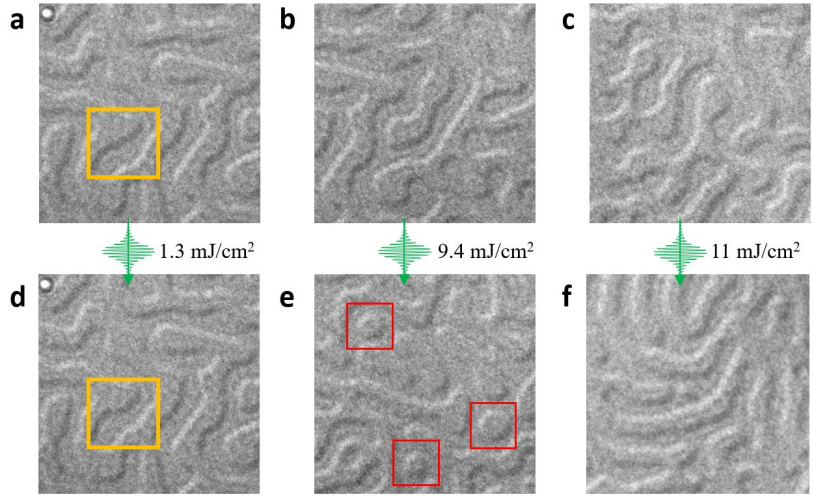

**Fig. S25. | Fluence dependent in-situ laser writing of skyrmions. a, b, c**. Ground states of the stripe domains obtained through zero-field cooling. **d, e, f.** Magnetic domain states after a single fs laser pulse with the fluence of 1.3 mJ/cm$^2$, 9.4 mJ/cm$^2$, and 11 mJ/cm$^2$, respectively. The red boxes indicated isolated skyrmions after a single fs laser pulse.

**Supplementary Note 9**: As depicted in Fig. S25a and S25d, under the condition of a single laser pulse fluence of 1.3 mJ/cm$^2$, the stripe domains within the yellow box merely exhibit domain wall movement after the laser pulse. In Fig. S25b and S25e, when we increase the single laser pulse fluence to 9.4 mJ/cm$^2$, the stripe domains become narrower and shorter, with some regions breaking to form skyrmions (highlighted in the red box). However, as we increase the single laser pulse fluence to 11 mJ/cm$^2$, stripe domains are formed without skyrmions (Fig. S25c and S25f). These in-situ laser fluence-dependent experiments indicate that a hybrid state with coexisting stripes and skyrmions is achievable without magnetic field.